\documentclass[a4paper,usenatbib,times,fleqn]{mn2e}
\usepackage{natbib,graphicx,amsmath,amssymb,times,txfonts}
\usepackage{aas_macros,url,pstricks,bm}

\def\rhog{\rho_{\mathrm{g}}}
\def\rhod{\rho_{\mathrm{d}}}
\def\ts{t_{\mathrm{s}}}

\def\deltav{\Delta \bm{v}}

\def\vb{\bm{v}}

\def\epsd{\epsilon}

\defcitealias{laibeprice11}{LP11}
\defcitealias{laibeprice12}{LP12a}
\defcitealias{laibeprice12a}{LP12b}
\defcitealias{laibeprice14}{LP14a}
\defcitealias{laibeprice14a}{LP14b}
\defcitealias{laibeprice14b}{LP14c}

\title[A fast algorithm for small grains in SPH]{A fast and explicit algorithm for simulating the dynamics of small dust grains with smoothed particle hydrodynamics}

\author[Price \& Laibe]{Daniel J. Price$^{1}$\thanks{daniel.price@monash.edu}, Guillaume Laibe$^{2}$\thanks{guillaume.laibe@gmail.com},  \\
$^{1}$Monash Centre for Astrophysics and School of Physics \& Astronomy, Monash University, Clayton, Vic 3800, Australia \\
$^{2}$School of Physics and Astronomy, University of St. Andrews, North Haugh, St. Andrews, Fife KY16 9SS, UK
}
\pagerange{\pageref{firstpage}--\pageref{lastpage}} \pubyear{2015}

\begin{document}
\label{firstpage}
\bibliographystyle{mn2e}
\maketitle

\begin{abstract}
We describe a simple method for simulating the dynamics of small grains in a dusty gas, relevant to micron-sized grains in the interstellar medium and grains of centimetre size and smaller in protoplanetary discs. The method involves solving one extra diffusion equation for the dust fraction in addition to the usual equations of hydrodynamics. This ``diffusion approximation for dust'' is valid when the dust stopping time is smaller than the computational timestep. We present a numerical implementation using Smoothed Particle Hydrodynamics (SPH) that is conservative, accurate and fast. It does not require any implicit timestepping and can be straightforwardly ported into existing 3D codes.
\end{abstract}


\section{Introduction}
\label{sec:intro}
 Small grains rule the interstellar medium (ISM). Micron-sized dust grains absorb ultraviolet radiation from hot, young stars and re-emit it in the infrared. Understanding how these grains interact with the gas is critical to understanding both the dynamics and thermodynamics of the ISM, and to interpreting observational results which usually assume a fixed gas-to-dust ratio in order to derive physical quantities such as the gas column density.
 
 Modelling such grains presents a severe computational challenge, since small grains are tightly coupled to the gas by the mutual drag force. This presents both a short timescale problem, since the stopping time of the grains is much shorter than the typical computational time, and a short lengthscale problem, since the physical separation between the dust grain population and the gas is much smaller than typical distances in the ISM.
 
  In a recent series of papers \citep{laibeprice12,laibeprice12a,laibeprice14,laibeprice14a,laibeprice14b} we have outlined the limitations associated with modelling dust and gas using the standard two fluid approach, where they are regarded as separate fluids coupled by a drag term. Typically the gas is represented by a set of particles or grid cells, while the dust is represented by a separate set of pressure-less particles coupled to the gas by a drag term. The length and timescale problems discussed above mean that with this approach one needs both infinite spatial and temporal resolution to accurately capture the dynamics of small grains in the limit of perfect coupling (\citealt{laibeprice12}; but see \citet{lorenbate14} for an alternative approach). However, this is the limit in which the mixture can be accurately described as a single fluid moving at the barycentric velocity. In \citet{laibeprice14,laibeprice14a} (hereafter \citetalias{laibeprice14,laibeprice14a}) we showed how the equations for a coupled dust-gas system can be reformulated to describe this single fluid mixture without loss of generality, solving both the length and timescale issues and also preventing artificial trapping of dust particles below the resolution of the gas (\citetalias{laibeprice12}; \citealt{ayliffeetal12}). The method is similar to the approach to other multi-fluid systems in astrophysics such as ionised plasmas \citep{pandeywardle08}, but more general since it can be implemented without any approximations.

 In \citetalias{laibeprice14a} we derived a Smoothed Particle Hydrodynamics (SPH) algorithm based on the fully general one fluid method and showed that it could accurately capture the dynamics of dust-gas mixtures in both the weakly coupled and tightly coupled limits. For problems involving small grains, however, the full machinery of the one fluid formulation is unnecessary and a much simpler and computationally inexpensive approach is possible, as outlined in Section 3.3 of \citetalias{laibeprice14}. This approximation is accurate when the stopping time, $t_{\rm s}$, is less than the Courant timestep (Eq. 115 in \citetalias{laibeprice14}). 
 
  Our goal in this paper is to derive a numerical implementation of this much simpler formulation, since there are many situations in astrophysics where the dynamics of small grains is the dominant effect. This includes simulations of galaxies, star formation in the interstellar medium --- where small grains control the thermodynamics --- and the settling and migration of dust in protoplanetary discs. We summarise the analytic formulation and its applicability in Sec.~\ref{sec:formulation}, the numerical implementation is described in Sec.~\ref{sec:numerics} and tests are presented in Sec.~\ref{sec:tests}. A public version of the \textsc{ndspmhd} code (v2.1) implementing the algorithms and with the precise setup of the test problems is released alongside this paper\footnote{\url{http://users.monash.edu.au/~dprice/ndspmhd/}}.

\section{The diffusion approximation for dust}
\label{sec:formulation}

\subsection{Continuum equations}

\subsubsection{General case}
In \citetalias{laibeprice14} we showed that, to first order in $\ts / T$, where $T$ is the timescale for a sound wave to propagate over a typical distance $L$, the equations describing the evolution of a dust-gas mixture can be written in the form 
\begin{eqnarray}
\frac{\mathrm{d} \rho}{\mathrm{d} t} & = & -\rho (\nabla\cdot\vb) \label{eq:cty},\\
\frac{\mathrm{d} \vb}{\mathrm{d}t}  & = & (1 - \epsilon)\bm{f}_{\rm g} + \epsilon \bm{f}_{\rm d} +\bm{f} \label{eq:mom}, \\
\frac{\mathrm{d}\epsd}{\mathrm{d}t} & = & -\frac{1}{\rho} \nabla\cdot\left[\epsd (1 - \epsd) \rho \ts \Delta \bm{f}  \right] \label{eq:eps}, \\
\frac{{\rm d} u}{{\rm d} t} & = &  -\frac{P}{\rhog} (\nabla\cdot\vb) + \epsilon \ts \left(  \Delta \bm{f} \cdot\nabla \right) u + \Lambda_{\rm heat} - \Lambda_{\rm cool}, \label{eq:dudt} \label{EQ:DUDT}
\end{eqnarray}
where $\rho$ is the total density of the mixture, $\epsd \equiv \rho_{\rm d}/\rho$ is the mass fraction of dust, $\bf f$ represents accelerations acting on both components of the fluid while $\bm{f}_{\rm g}$ and $\bm{f}_{\rm d}$ represent the accelerations acting on the gas and dust components, respectively, $\Delta\bm{f} \equiv \bm{f}_{\rm d} - \bm{f}_{\rm g}$ is the differential acceleration between the gas and dust, $u$ is the specific thermal energy of the gas, $P$ is the gas pressure, and $\Lambda_{\rm heat}$ and $\Lambda_{\rm cool}$ are additional heating and cooling terms, respectively\footnote{Eq.~\ref{eq:dudt} differs from the expression we gave for the ``first order approximation'' in \citetalias{laibeprice14}. The drag heating term, $\epsilon \deltav^{2}/\ts$, is clearly negligible in the terminal velocity approximation and the $P{\rm d}V$ work term should involve $\nabla\cdot\bm{v}$ rather than $\nabla\cdot\bm{v}_{\rm g}$. Both approximations are required for the numerical scheme to conserve total energy as defined in the terminal velocity approximation (Eq.~\ref{eq:energy}).}
. The velocity $\bf v$ is the barycentric velocity of the mixture, defined as
\begin{equation}
\bm{v} \equiv \frac{\rho_{\rm d}\bm{v}_{\rm d} + \rho_{\rm g}\bm{v}_{\rm g}}{\rho} = \epsd \bm{v}_{\rm d} + (1 - \epsd) \bm{v}_{\rm g},
\end{equation}
In the so-called \emph{terminal velocity approximation} \citep{youdingoodman05,chiang08,barranco09,leeetal10,jbl11} assumed in Equations \ref{eq:cty}--\ref{eq:dudt}, $\Delta\bm{f}$ is rapidly balanced by the drag. Thus, the time dependence of the differential velocity can be ignored, and the differential velocity between the gas and dust is given by
\begin{equation}
\Delta \bm{v} \equiv (\bm{v}_{\rm d} - \bm{v}_{\rm g}) \simeq \ts \Delta \bm{f}.
\label{eq:deltavtv}
\end{equation}
This also implies that the anisotropic pressure term in the momentum equation \citepalias[see][]{laibeprice14} should be neglected. The terminal velocity approximation is valid when the drag coefficient $K$ is large such that the stopping time,
\begin{equation}
\ts \equiv \frac{\rho_{\rm d} \rho_{\rm g}}{K (\rho_{\rm d} + \rho_{\rm g})} = \frac{\epsd (1 - \epsd) \rho}{K},
\end{equation}
is short compared to the timestep. Various physical prescriptions for $K$ in the Epstein and Stokes drag regimes are given in \citet{laibeprice12a} but the essential point is that $K$ is inversely proportional to the grain size, being large for small grains.

The differential acceleration $\Delta\bm{f}$ depends on the physics in the problem, i.e. the forces affecting the gas but not the dust, which may include pressure, magnetic and other forces. In our numerical implementation we consider the contributions from the pressure gradient (see below) and also the artificial viscosity term, which should likewise affect the gas only.

\subsubsection{Hydrodynamics}
\label{sec:hydro}
For the simple case of hydrodynamics, the only force is the pressure gradient, giving
\begin{equation}
\bm{f}_{\rm g} = -\frac{\nabla P}{\rho_{g}}; \hspace{5mm} \bm{f}_{\rm d} = 0,
\end{equation}
and thus
\begin{equation}
\Delta \bm{f} = \frac{\nabla P}{\rho_{g}},
\label{eq:df}
\end{equation}
giving Equations \ref{eq:cty}--\ref{eq:dudt} in the form
\begin{eqnarray}
\frac{\mathrm{d} \rho}{\mathrm{d} t} & = & -\rho (\nabla\cdot\vb) \label{eq:ctyalt},\\
\frac{\mathrm{d} \vb}{\mathrm{d}t}  & = & -\frac{\nabla P}{\rho}  + \bm{f} \label{eq:momalt}, \\
\frac{\mathrm{d}\epsd}{\mathrm{d}t} & = & -\frac{1}{\rho} \nabla\cdot\left(\epsilon \ts \nabla P  \right), \label{eq:epsalt} \\
\frac{{\rm d} u}{{\rm d} t} & = &  -\frac{P}{\rhog} (\nabla\cdot\vb) + \frac{\epsilon \ts}{\rhog} \left(  \nabla P \cdot\nabla u \right) + \Lambda_{\rm heat} - \Lambda_{\rm cool}. \label{eq:dudtalt}
\end{eqnarray}
These are similar to the usual equations of hydrodynamics in the absence of dust. The only differences are the extra equation that describes the evolution of the dust fraction; the modifications to the thermal energy equation; and the fact that the pressure is related to the \emph{gas} density only, not the total density (see Sec.~\ref{sec:eos} below; this gives the zeroth order effect of a `heavy fluid', as discussed in \citetalias{laibeprice14}).

\subsubsection{Equation of state}
\label{sec:eos}
The equation set is closed by the usual equation of state specifying the gas pressure $P$ in terms of the gas density and temperature. Unless otherwise specified in this paper we assume an adiabatic equation of state, i.e.
\begin{equation}
P = (\gamma - 1) \rho_{\rm g} u = (\gamma - 1)(1 - \epsilon) \rho u,
\label{eq:eos}
\end{equation}
where $\gamma$ is the usual adiabatic constant.

\subsection{Timestepping}
\label{sec:timestepping}
 The main change when adopting the formulation given above compared to hydrodynamics is the addition of the diffusion equation for the dust fraction (\ref{eq:epsalt}). This introduces an additional constraint on the timestep when the diffusion coefficient is large. Assuming an isothermal equation of state $P = c_{\rm s}^{2} \rho_{\rm g} = c_{\rm s}^{2} (1- \epsilon)\rho$ and a constant density, (\ref{eq:epsalt}) can be written as a simple diffusion equation for $\epsilon$
\begin{equation}
\frac{\mathrm{d}\epsd}{\mathrm{d}t} = \nabla\cdot\left(\eta \nabla \epsilon  \right),
\label{eq:diffuse}
\end{equation}
where the diffusion coefficient $\eta \equiv \epsilon \ts c_{s}^{2}$. This implies a stability constraint of the form
\begin{equation}
\Delta t < \Delta t_{\epsilon} = C_{0} \frac{h^{2}}{\eta} = C_{0} \frac{h^{2}}{\epsilon t_{\rm s} c_{\rm s}^{2} }, \label{eq:dtd}
\end{equation}
where $C_{0}$ is a dimensionless safety factor of order unity and $h$ is the resolution length (the smoothing length in SPH). We can rewrite (\ref{eq:dtd}) as
\begin{equation}
\Delta t < C \left(\frac{ \Delta t_{\rm Cour}}{t_{\rm s}}\right)^{2}  \ts ,
\end{equation}
where $C$ is a constant and $ \Delta t_{\rm Cour} = C_{0} h / c_{\rm s}$ is the usual Courant condition. This implies that the timestep is constrained when the stopping time is \emph{long} --- the opposite of the usual situation where the timestep is constrained when the stopping time is short. This is the main advantage of using the diffusion approximation --- small grains can be integrated explicitly.

Specifically, the diffusion timestep becomes the limiting timestep when
\begin{equation}
\epsilon \ts > \Delta t_{\rm Cour}.
\end{equation}
However, this is also the criterion for when the terminal velocity approximation breaks down \citepalias[see][]{laibeprice14}. This implies that the diffusion approximation becomes inaccurate precisely when the timestep implied by (\ref{eq:dtd}) starts to constrain the timestep, because at this point the time-dependence in $\Delta \bm{v}$ becomes important. Once this occurs, one should revert to the general formulation given by \citetalias{laibeprice14a} where $\Delta \bm{v}$ is explicitly evolved, or a two-fluid method. Physically this transition occurs once grains grow beyond a certain size, implying that the stopping time becomes long, or equivalently when one has enough temporal resolution to resolve the timescale on which the differential velocity is changing. 


\subsection{Validity of the diffusion approximation for astrophysics}
\label{sec:validity}
Under what circumstances is the diffusion approximation valid for astrophysics? Consider a drag force described by the linear Epstein regime, appropriate to small grains at low Mach number. In this case the drag coefficient is given by \citepalias[e.g.][]{laibeprice12a}
\begin{equation}
K = \rho_{\rm g} \rho_{\rm d} \frac{4\pi}{3} \frac{s_{\rm grain}^{2}}{m_{\rm grain}} \sqrt{\frac{8}{\pi\gamma}} c_{\rm s},
\end{equation}
where $s_{\rm grain}$ is the grain size and $m_{\rm grain}$ is the mass of an individual grain. Assuming $m_{\rm grain} = \frac43 \pi \rho_{\rm grain} s_{\rm grain}^{3}$, where $\rho_{\rm grain}$ is the intrinsic grain density, the stopping time is
\begin{equation}
t_{\rm s} = \frac{ \rho_{\rm grain} s_{\rm grain}}{\rho c_{\rm s}} \sqrt{\frac{\pi\gamma}{8}}.
\label{eq:tseps}
\end{equation}

\subsubsection{Grains in the interstellar medium}
Evaluating this for dust grains in a molecular cloud, we have
\begin{equation}
t_{\rm s} = 2.5 \times 10^{3} {\rm yr} \left(\frac{\rho_{\rm grain}}{1 {\rm g}/{\rm cm}^{-3}}\right) 
\left(\frac{s_{\rm grain}}{0.1 \micron} \right)
\left( \frac{\rho}{10^{-20}  {\rm g}/{\rm cm}^{3}} \right)^{-1}
\left(\frac{c_{\rm s}}{0.2 {\rm km/s}} \right)^{-1}.
\end{equation}
This indicates that the diffusion approximation is valid for small grains in the interstellar medium, since the stopping time is much smaller than the dynamical time ($\sim 10^{6} {\rm yr}$).

\subsubsection{Protoplanetary discs}
 For a protoplanetary disc, the relevant comparison is to the orbital timescale since the pressure timescale $H/c_{\rm s} \equiv 1/\Omega$. A reasonable criterion for validity is therefore that
\begin{equation}
t_{\rm s} \Omega \approx \frac{\rho_{\rm grain} s_{\rm grain}}{\Sigma} \ll 1.
\end{equation}
This suggests the approximation is valid for grain sizes
\begin{equation}
s_{\rm grain} \ll 10^{2} {\rm cm} \left(\frac{\Sigma}{10^{2} {\rm g}/{\rm cm}^{2}}\right)  \left(\frac{\rho_{\rm grain}}{1 {\rm g}/{\rm cm}^{3}}\right)^{-1}.
\end{equation}
Hence diffusion is a reasonable approximation for grains of $\sim$cm size and smaller in protoplanetary discs. This maximum size is smaller in the outer disc regions, since typically the surface density is inversely proportional to distance from the central star. We examine this experimentally in Section~\ref{sec:settle}.


\section{Implementation in Smoothed Particle Hydrodynamics}
\label{sec:numerics}

\subsection{Implementation using two first derivatives}
\label{sec:firstderivs}
The SPH representation of a more general form of Eqs.~\ref{eq:cty}--\ref{eq:dudt} have been derived in \citetalias{laibeprice14a} and so our first approach is to adopt the same discretisation but with $\Delta \bm{v}$ prescribed by Eq.~\ref{eq:deltavtv}, giving
\begin{eqnarray}
\rho_{a} & = & \sum_{b} m_{b} W_{ab} (h_{a}), \label{eq:sphcty} \\
\frac{{\rm d}\bm{v}_a}{{\rm d}t} & = & -\sum_{b} m_{b} \left[ \frac{P_{a} + q^{AV}_{ab, a}}{\Omega_{a} \rho_{a}^{2}} \nabla_{a} W_{ab} (h_{a}) + \frac{P_{b} + q^{AV}_{ab, b}}{\Omega_{b} \rho_{b}^{2}} \nabla_{a} W_{ab} (h_{b}) \right] \nonumber \\
& & + \bm{f}_{a}, \label{eq:sphmom} \\
\frac{{\rm d}\epsilon_a}{{\rm d}t} & = & -\sum_{b} m_{b} \left[\frac{\epsilon_{a}(1 - \epsilon_{a}) t_{{\rm s}, a}}{\Omega_{a}\rho_{a}} \Delta \bm{f}_{a}\cdot \nabla_{a} W_{ab} (h_{a})\right. \nonumber \\
& & \phantom{-\sum_{b} m_{b}} + \left.\frac{\epsilon_{b}(1 - \epsilon_{b})  t_{{\rm s}, b}}{\Omega_{b}\rho_{b}} \Delta \bm{f}_{b}\cdot \nabla_{a} W_{ab} (h_{b}) \right] \label{eq:spheps}, \\
\frac{{\rm d}u_{a}}{{\rm d}t} & = & \frac{1}{\Omega_{a} (1 - \epsilon_{a})\rho_{a}^{2}} \sum_{b} m_{b} (P_{a} + q^{AV}_{ab, a}) \left( \bm{v}_{a} - \bm{v}_{b}\right)\cdot\nabla_{a} W_{ab} (h_{a}) \nonumber \\
&& - \frac{\epsilon_{a} t_{{\rm s}, a}}{\Omega_{a}\rho_{a}}  \Delta \bm{f}_{a} \cdot \sum_{b} m_{b} (u_{a} - u_{b}) \nabla_{a} W_{ab} (h_{a}), \label{eq:sphdudt}
\end{eqnarray}
where $W_{ab}$ is the usual SPH kernel (we use the usual cubic spline kernel throughout this paper unless otherwise indicated), $h$ is the smoothing length, $\Omega$ is the usual term related to smoothing length gradients
\begin{equation}
\Omega_{a} \equiv 1 - \frac{\partial h_{a}}{\partial \rho_{a}} \sum_{b} m_{b} \frac{\partial W_{ab}(h_{a})}{\partial h_{a}},
\end{equation}
and $h$ is related to $\rho$ in the usual manner requiring an iterative procedure to solve Eq.~\ref{eq:sphcty} (\citetalias{laibeprice14a}; \citealt{pricemonaghan04a,pricemonaghan07}) and unless otherwise specified we use a ratio of $h$ to particle spacing of 1.2 \citep{price12}.
The reader will notice that the first two equations are identical to the usual density summation and momentum equation in SPH. The only differences, mirroring the continuum case (Eqs.~\ref{eq:ctyalt}--\ref{eq:dudtalt}), are the addition of the diffusion equation (\ref{eq:eps}) for the dust fraction, the extra terms in the thermal energy equation (\ref{eq:sphdudt}) and the dependence of the pressure on the gas density rather than the total density in the equation of state (\ref{eq:eos}).

 The differential force between the fluids implied by our formulation of Eq.~\ref{eq:sphmom} is
\begin{equation}
\Delta \bm{f}_{a} = -\bm{f}^{a}_{\rm g},
\end{equation}
where
\begin{equation}
(1 - \epsilon_{a}) \bm{f}^{a}_{\rm g} = -\sum_{b} m_{b} \left[ \frac{P_{a} + q^{AV}_{ab, a}}{\Omega_{a} \rho_{a}^{2}} \nabla_{a} W_{ab} (h_{a}) + \frac{P_{b} + q^{AV}_{ab, b}}{\Omega_{b} \rho_{b}^{2}} \nabla_{a} W_{ab} (h_{b}) \right]. \label{eq:deltafg}
\end{equation}
This $\Delta \bm{f}$, computed as above, is then used to evaluate Equations~\ref{eq:spheps} and \ref{eq:sphdudt}, requiring a separate loop over the particles.

\subsection{Shock-capturing terms}
\label{sec:av}

\subsubsection{Artificial viscosity}
We formulate the artificial viscosity term following the more general algorithm derived in \citetalias{laibeprice14a} but slightly modified to appear as separate $q_{a}$ and $q_{b}$ terms to avoid averaging the kernel gradients, following the formulation of artificial viscosity used in the \textsc{Phantom} code \citep{pricefederrath10,lodatoprice10}. We use
\begin{equation}
q^{AV}_{ab, a} = \begin{cases}
-\frac12 \rho_{a} (1 - \epsilon_{a}) v_{{\rm sig}, a} \bm{v}_{ab} \cdot \hat{\bf r}_{ab}.
&  \bm{v}_{ab} \cdot \hat{\bf r}_{ab} < 0 \\
0 & \bm{v}_{ab} \cdot \hat{\bf r}_{ab} \geq 0
\end{cases}
\label{eq:qterm}
\end{equation}
where $\bm{v}_{ab} \equiv \bm{v}_{a} - \bm{v}_{b}$ (similarly for ${\bf r}_{ab}$) and the signal speed $v_{\rm sig}$ corresponds to the usual choice for hydrodynamics, i.e.
\begin{equation}
v_{{\rm sig},a} = \alpha_{a} c_{{\rm s}, a} + \beta \vert \bm{v}_{ab} \cdot {\bf r}_{ab}\vert.
\label{eq:vsig}
\end{equation}
where $\alpha \in [0,1]$ is the linear dimensionless viscosity parameter (in general this can be individual to each particle, e.g. when using the \citealt{morrismonaghan97} or \citealt{cullendehnen10} switches) and $\beta$ (typically $\beta=2$) is the Von Neumann-Richtmeyer viscosity parameter.

 The $q^{AV}$ term and the signal speed involve the jump in total velocity rather than the gas velocity, unlike in \citetalias{laibeprice14a} where only the gas velocity is used. This is both physical and practical: In the terminal velocity approximation the difference
\begin{equation}
\bm{v} - \bm{v}_{\rm g} \equiv \epsilon \ts {\Delta \bm{f}},
\end{equation}
is small by definition. The practical side is that it we do not know $\Delta \bm{f}$ prior to the evaluation of Eq.~\ref{eq:sphmom}, so it is not possible to use the gas velocity directly in the artificial viscosity term without an iterative approach.

\subsubsection{Artificial conductivity}
We write the artificial conductivity term, necessary for correct treatment of contact discontinuities \citep{price08}, similar to that in \citetalias{laibeprice14a}, giving
\begin{equation}
\left(\frac{{\rm d}u_{a}}{{\rm d}t}\right)_{\rm cond} = \frac{1}{1 - \epsilon_{a}} \sum_{b} m_{b} \left[ \frac{Q_{ab, a}}{\Omega_{a} \rho_{a}^{2}}F_{ab} (h_{a}) + \frac{Q_{ab, b}}{\Omega_{b} \rho_{b}^{2}}F_{ab} (h_{b})  \right],
\end{equation}
where $\nabla_{a} W_{ab} \equiv F_{ab} \hat{\bf r}_{ab}$ and
\begin{equation}
Q_{ab, a} =  \frac12 \alpha_{u} \rho_{a} v_{\rm sig, u}  (u_{a} - u_{b}),
\end{equation}
with $\alpha_{u} \in [0,1]$ the dimensionless conductivity parameter and $v_{\rm sig, u} = \vert \bm{v}_{ab}\cdot\hat{\bf r}_{ab} \vert$ \citep{price08,wvc08}.

\subsection{Conservation properties}
 Equation~\ref{eq:sphcty} manifestly conserves the total mass since the mass of the SPH particles is constant. Similarly it can be straightforwardly verified that the total momentum is conserved, since
\begin{equation}
\frac{{\rm d}}{{\rm d}t} \sum_{a} m_{a} \bm{v}_{a} = \sum_{a} m_{a} \frac{{\rm d}\bm{v}_a}{{\rm d}t} = 0,
\end{equation}
due to the fact that the resulting double summation is antisymmetric in the particle indices $a$ and $b$. Likewise the total angular momentum is conserved, since
\begin{equation}
\frac{{\rm d}}{{\rm d}t} \sum_{a} m_{a} {\bf r}_{a} \times \bm{v}_{a} = \sum_{a} m_{a} {\bf r}_{a} \times \frac{{\rm d}\bm{v}_a}{{\rm d}t} = 0.
\end{equation}
(for more details, see Equation 33 in \citealt{price12}). Finally, one may also verify that the total mass of each species is conserved, since
\begin{equation}
\frac{{\rm d} M_{\rm d}}{{\rm d}t} = -\frac{{\rm d} M_{\rm g}}{{\rm d}t} = \sum_{a} m_{a} \frac{{\rm d}\epsilon_{a}}{{\rm d} t} = 0.
\label{eq:dmgasdt}
\end{equation}
The proof is identical to that given in \citetalias{laibeprice14a} and again results from the fact that the double summation is antisymmetric with respect to the particle indices.
 
 The total energy of the mixture in the terminal velocity approximation is given by \citepalias{laibeprice14}
\begin{equation}
E = \int \left( \frac12 \rho \bm{v}^{2} + \rhog u \right) {\rm d}V = \int \left[ \frac12 \rho \bm{v}^{2} + \rho(1 - \epsilon) u \right] {\rm d}V.
\label{eq:energy}
\end{equation}
This is simpler than the full one fluid expression (Eq.~61 in \citetalias{laibeprice14}) as the term involving $\deltav^{2}$ can be neglected.
Discretised onto the mixture particles, the energy becomes
\begin{equation}
E = \sum_{a} m_{a} \left[ \frac12 \bm{v}_{a}^{2} + (1 - \epsilon_{a}) u_{a} \right],
\label{eq:energySPH}
\end{equation}
 Conservation of energy implies that
\begin{equation}
\frac{{\rm d}E}{{\rm d}t} = \sum_{a} m_{a} \left[ \bm{v}_{a}\cdot\frac{{\rm d}\bm{v}_{a}}{{\rm d}t} + (1 - \epsilon_{a}) \frac{{\rm d}u_{a}}{{\rm d}t}  - u_{a} \frac{{\rm d}\epsilon_{a}}{{\rm d}t} \right] = 0.
\end{equation}
Substituting Equations \ref{eq:sphmom} and \ref{eq:spheps} in the above, we require for energy conservation that
\begin{align}
\sum_{a} m_{a} (1 - \epsilon_{a}) \frac{{\rm d}u_{a}}{{\rm d}t} & = \sum_{a} \sum_{b} m_{a} m_{b} \left[ \frac{P_{a} + q^{AV}_{ab,a}}{\Omega_{a}\rho_{a}^{2}} \bm{v}_{a}\cdot \nabla_{a} W_{ab} (h_{a})\right] \nonumber\\
& + \sum_{a} \sum_{b} m_{a} m_{b} \left[ \frac{P_{b} + q^{AV}_{ab,b}}{\Omega_{b}\rho_{b}^{2}} \bm{v}_{a}\cdot \nabla_{a} W_{ab} (h_{b})\right] \nonumber \\
& - \sum_{a} \sum_{b} m_{a} m_{b} \left[ \frac{u_{a}(1-\epsilon_{a})\epsilon_{a}t_{{\rm s}, a}}{\Omega_{a}\rho_{a}} \Delta \bm{f}_{a}\cdot \nabla_{a} W_{ab} (h_{a})\right] \nonumber \\
& - \sum_{a} \sum_{b} m_{a} m_{b} \left[ \frac{u_{a}(1-\epsilon_{b})\epsilon_{b}t_{{\rm s}, b}}{\Omega_{b}\rho_{b}} \Delta \bm{f}_{b}\cdot \nabla_{a} W_{ab} (h_{b})\right]. \nonumber
\end{align}
Swapping the summation indices $a$ and $b$ in the second and fourth terms, using the antisymmetry of the kernel gradient $\nabla_{b} W_{ba}(h_{a}) = -\nabla_{a} W_{ab} (h_{a})$ and collecting terms we have
\begin{align}
\sum_{a} m_{a} (1 - \epsilon_{a}) \frac{{\rm d}u_{a}}{{\rm d}t}  & = \sum_{a} \sum_{b} m_{a} m_{b} \left[ \frac{P_{a} + q^{AV}_{ab,a}}{\Omega_{a}\rho_{a}^{2}} (\bm{v}_{a} - \bm{v}_{b})\cdot \nabla_{a} W_{ab} (h_{a})\right] \nonumber\\
& - \sum_{a} \sum_{b} m_{a} m_{b} \left[ \frac{(1-\epsilon_{a})\epsilon_{a}t_{{\rm s}, a}}{\Omega_{a}\rho_{a}} (u_{a} - u_{b})\Delta \bm{f}_{a}\cdot \nabla_{a} W_{ab} (h_{a})\right], \nonumber \\
\end{align}
from which it is straightforward to verify that, with ${\rm d}u_{a}/{\rm d}t$ given by Eq.~\ref{eq:sphdudt}, total energy is conserved exactly.

 Thus, the approximate version of the one fluid algorithm retains all of the conservation properties of both the original SPH method and the general one fluid approach derived in \citetalias{laibeprice14a}.

\subsection{Implementation using direct second derivatives}
\label{sec:2ndderiv}
The main disadvantage of the formulation given above is that it requires a third loop over the particles to compute the ${\rm d}\epsilon/{\rm d}t$ term, beyond the two loops required for the density and force, respectively. This is because $\Delta \bm{f}$ is required before Eq.~\ref{eq:spheps} can be evaluated, but must be computed after the right hand side of (\ref{eq:sphmom}) is known. Thus in general this scheme is 1/3 more expensive than a standard SPH code. Here we provide an alternative scheme that does not require this extra loop. The two implementations are compared in Section~\ref{sec:tests}.

\subsubsection{Diffusion equation for the dust fraction}
 We can avoid the extra loop over the particles by discretising the second derivative in Eq.~\ref{eq:eps} directly, similar to the usual way that dissipative terms are treated in SPH. To do this we assume that viscous forces do not significantly drive the differential velocity between the fluids, i.e. that $\Delta \bm{f}$ is given by Eq.~\ref{eq:df} and therefore that Eq.~\ref{eq:eps} is given by Eq.~\ref{eq:epsalt}. We then discretise Eq.~\ref{eq:epsalt} in the usual manner following \citet{clearymonaghan99}:
\begin{equation}
\frac{\mathrm{d}\epsilon_{a}}{\mathrm{d}t} = - \sum_{b} \frac{m_{b}}{\rho_{a}\rho_{b}} (D_{a} + D_{b}) \left(P_{a} - P_{b} \right) \frac{\overline{F}_{ab}}{\vert r_{ab} \vert},
\label{eq:depsdtsph}
\end{equation}
where $D \equiv \epsilon \ts$, $\overline{F}_{ab} \equiv \frac12 [F_{ab} (h_{a}) + F_{ab} (h_{b})]$ and $F_{ab}$ is defined such that $\nabla W_{ab} \equiv F_{ab} \hat{\bf r}_{ab}$. It is straightforward to show that this expression also conserves both the total mass of dust and gas, since the resulting double summation in Eq.~\ref{eq:dmgasdt} is antisymmetric with respect to the particle index.

\subsubsection{Harmonic vs. arithmetic mean}
 In the original \citet{clearymonaghan99} paper (see also \citealt{monaghan05}) it was suggested to use the harmonic mean instead of the arithmetic mean of the diffusion coefficient, i.e.
\begin{equation}
\frac{\mathrm{d}\epsilon_{a}}{\mathrm{d}t} = - \sum_{b} \frac{m_{b}}{\rho_{a}\rho_{b}} \frac{4 D_{a}D_{b}}{(D_{a} + D_{b})} \left(P_{a} - P_{b} \right) \frac{\overline{F}_{ab}}{\vert r_{ab} \vert},
\label{eq:depsdtsphcm}
\end{equation}
with the motivation being that this better handles the case where the diffusion coefficient $D$ is discontinuous. However, we found this could give incorrect results. Imagine the dust confined to a layer such that $\epsilon_{a} = 0$ for some particle, a, outside the layer, with $\epsilon_{b} \neq 0$ for particles inside the layer. In this case the harmonic mean is zero for every pair involving particle $a$ since ${\rm d}\epsilon_{a}/{\rm d}t$ is \emph{always} zero. Thus it is impossible for the layer to move into the region where $\epsilon$ was initially zero, which is clearly incorrect (consider for example a discrete layer of dust descending under gravity). With the arithmetic mean we find no such problem and it is easy to prove that the formulation is correct\footnote{While \citet{clearymonaghan99} proposed the harmonic mean, there is no detailed comparison between the two choices in their paper and the only proof that the harmonic mean correctly represents the second derivative, apart from the numerical tests in their paper, involves a Taylor-series approximation where the harmonic mean reduces to the arithmetic mean.}, for example with a procedure similar to the one we use in Appendix~\ref{sec:dudtalt}.

\subsubsection{Thermal energy equation}
 In order to conserve energy, the corresponding expression for ${\rm d}u/{\rm d}t$ when using Equation~\ref{eq:depsdtsph} for ${\rm d}\epsilon/{\rm d}t$ is given by
\begin{eqnarray}
\frac{{\rm d}u_{a}}{{\rm d}t} & = & \frac{1}{\Omega_{a} (1 - \epsilon_{a})\rho_{a}^{2}} \sum_{b} m_{b} (P_{a} + q^{AV}_{ab, a}) \left( \bm{v}_{a} - \bm{v}_{b}\right)\cdot\nabla_{a} W_{ab} (h_{a}) \nonumber \\
& - & \frac{1}{2(1-\epsilon_{a})\rho_{a}}\sum_{b} \frac{m_{b}}{\rho_{b}} (u_{a} - u_{b}) (D_{a} + D_{b}) (P_{a} - P_{b}) \frac{\overline{F}_{ab}}{\vert r_{ab}\vert}, \label{EQ:SPHDUDTALT}
\end{eqnarray}
At first sight the second term is a rather strange one and it is not at all clear that this should translate to the correct physical term in Eq.~\ref{eq:sphdudt}. Yet, amazingly, it does --- the proof is given in Appendix~\ref{sec:dudtalt}. Hence there is no disadvantage in using this alternative formulation with respect to conservation properties. The shock capturing terms remain the same as in Section~\ref{sec:av}.

\subsubsection{Choice of smoothing kernel}
Although the formulation of second derivatives in SPH using the kernel gradient (\ref{eq:depsdtsph}) is now more than 30 years old \citep{brookshaw85}, and while it is clearly better than using $\nabla^{2} W$ directly, to our knowledge there has been no systematic investigation of the best kernel to use in order to compute a second derivative. In particular, on the dust settling test in Section~\ref{sec:settle} we found that using (\ref{eq:depsdtsph}) with the cubic spline could give quite noisy results. Hence for this test we instead adopted the $M_{6}$ quintic kernel instead (see Sec.~\ref{sec:settle}). While this results in a more accurate estimate, it is also more expensive due to the larger kernel radius. Hence a more systematic investigation of suitable kernels for second derivatives in SPH would be valuable here. For example, in \citetalias{laibeprice12} we found double-hump shaped kernels to be an order of magnitude more accurate compared to standard kernels for computing the drag terms in the two fluid method at no additional cost.

\subsubsection{Two first derivatives vs. direct second derivatives}
To our knowledge there exists no systematic study on whether it is better to compute second derivatives in SPH directly or using two consecutive first derivatives (though see \citealt{watkinsetal96}). In principle both approaches yield a second order approximation provided that the particles are well ordered, and in the context of implementing physical viscosity terms in SPH both approaches have been advocated \citep[e.g.][]{flebbeetal94,watkinsetal96,espanolrevenga03,lodatoprice10}, with only \citet{watkinsetal96} suggesting that the two first derivatives approach is more accurate. By comparing our two implementations in Sec.~\ref{sec:tests} we effectively compare both approaches. We find only small differences between the two approaches in terms of the overall accuracy, with the main advantages being that the direct second derivatives approach is both faster and easier to implement.

\section{Numerical tests}
\label{sec:tests}
 A key issue in developing numerical codes for dust-gas mixtures is that there are few simple test problems that can be used to  benchmark the algorithm. We have partially resolved this issue by deriving the analytic solution for linear waves in such a mixture \citep{laibeprice11} and showing that the solution for a shock in the limit where $\Delta \bm{v} \to 0$ is the same as for the hydrodynamic case but with a modified sound speed \citep{laibeprice12, miuraglass82}. The \textsc{dustybox} solution \citep{laibeprice11} is not relevant to this paper since we have already assumed that $\Delta \bm{v}$ has reached its asymptotic value by using the terminal velocity approximation. Hence, we use the \textsc{dustywave} and \textsc{dustyshock} problems to benchmark our algorithm. Our exploration of the diffusion approximation for dust suggested a new test problem with a simple analytic solution, which we describe in Sec.~\ref{sec:dustydiffusion}.

\begin{figure*} 
   \centering
   \includegraphics[width=\columnwidth]{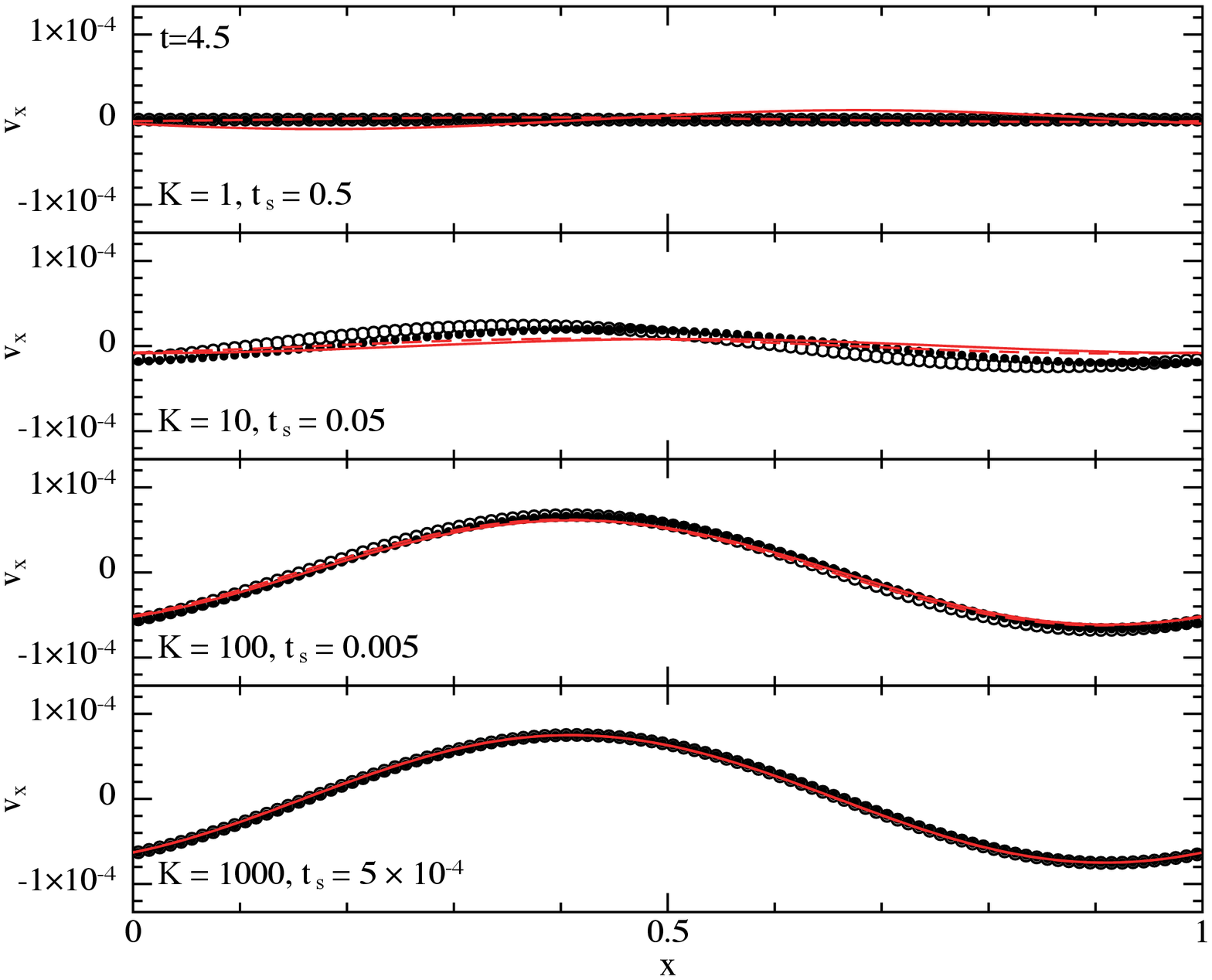}
   \includegraphics[width=\columnwidth]{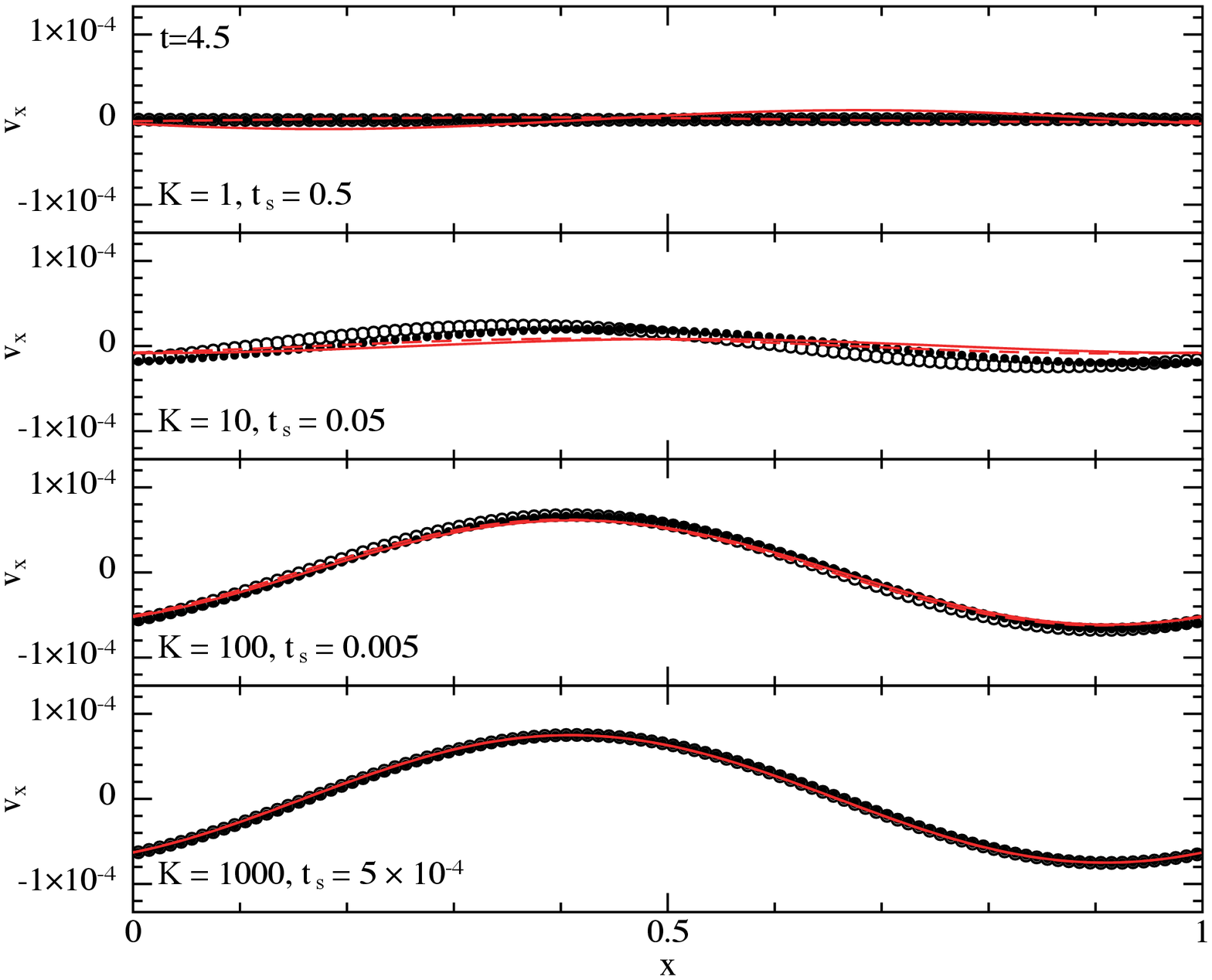}
   \caption{Gas and dust velocities (filled and open circles, respectively) in the dustywave problem using 100 SPH particles and our two implementations of the dust diffusion approximation: Two first derivatives (left) and direct second derivatives (right). These may be compared to the analytic linear solution from \citetalias{laibeprice11} given by the red solid (gas) and dashed (dust) lines. The $L_{2}$ error is within 6\% of the analytic solution for $K=100$ and within 2\% for $K=1000$, where the diffusion approximation is applicable (here for $K \gtrsim 42$ corresponding to $\ts > \Delta t_{\rm Cour} = 0.012$). The solution becomes inaccurate at weaker drag ($\ts > 0.012$). There is no discernible difference between the two implementations, except that the implementation with direct second derivatives (right) is faster.}
   \label{fig:dustywave-vels}
\end{figure*}


\subsection{\sc Dustywave}
 In \textsc{dustywave} problem, we solve for the propagation of a linear wave in a dust-gas mixture. We set up the problem in 1D as in our previous papers \citepalias{laibeprice12,laibeprice12a,laibeprice14a}, using $\rho_{{\rm d},0} = \rho_{{\rm g},0} = 1$ (i.e. $\rho_{0} = 2$ and $\epsilon_{0} = 0.5$) with a sinusoidal perturbation to the velocity and density of the mixture particles $v(x) = v_{0} \sin (2\pi x)$ and $\rho(x) = \rho_{0}\left[1 + \delta\rho_{0} \sin(2\pi x)\right]$, with amplitude $v_{0} = \delta\rho_{0} = 1 \times 10^{-4}$, with a corresponding thermal energy perturbation given by $\delta u = P_{0} / \rho_{{\rm g}, 0}^{2} \delta \rho_{\rm g}$. An adiabatic equation of state is used with $\gamma = 5/3$ and the thermal energy is set so that the initial sound speed $c_{\rm s} = 1$. We use 100 SPH particles in the domain $x \in [0,1]$.

\begin{figure*} 
   \centering
   \includegraphics[width=\columnwidth]{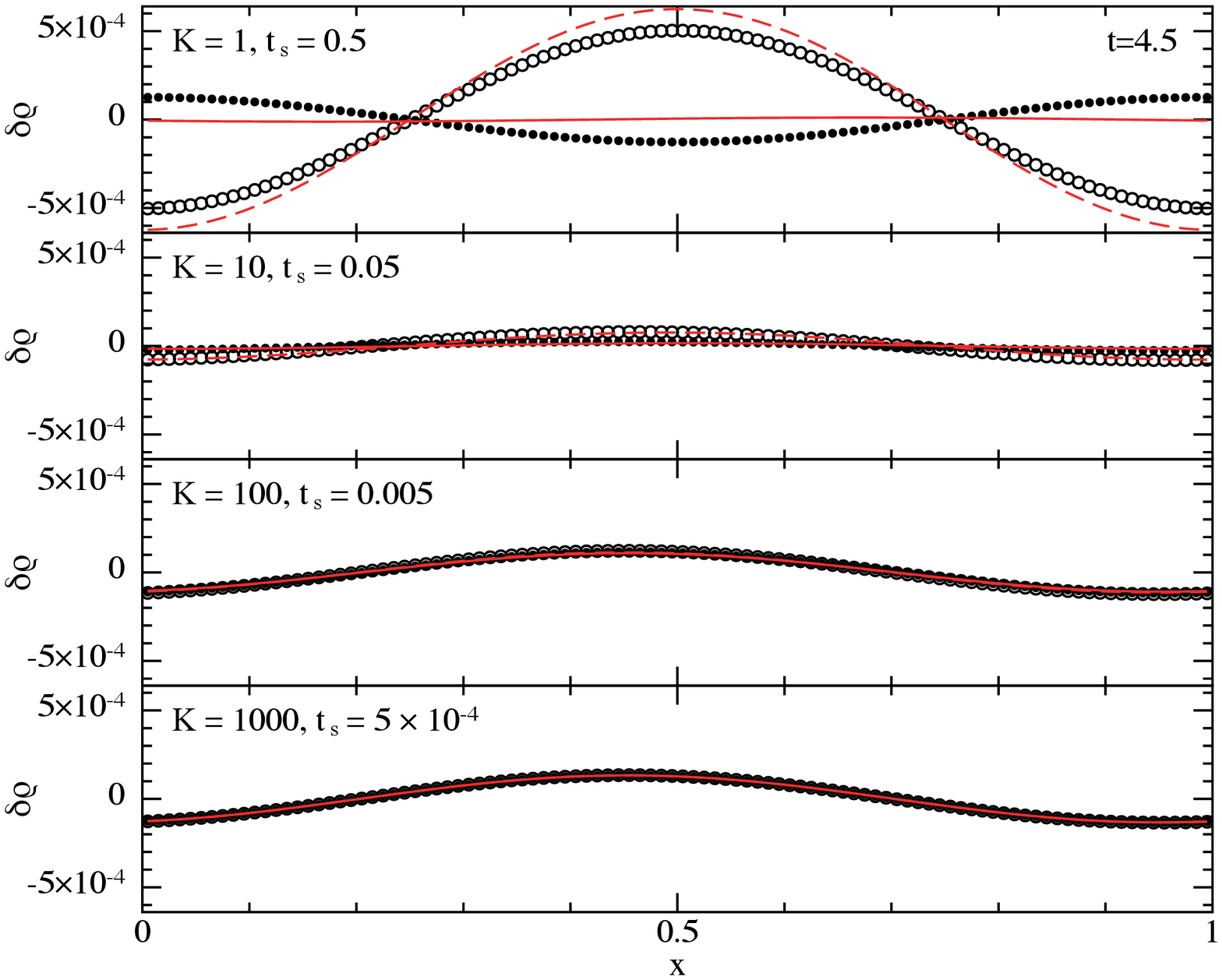}
   \includegraphics[width=\columnwidth]{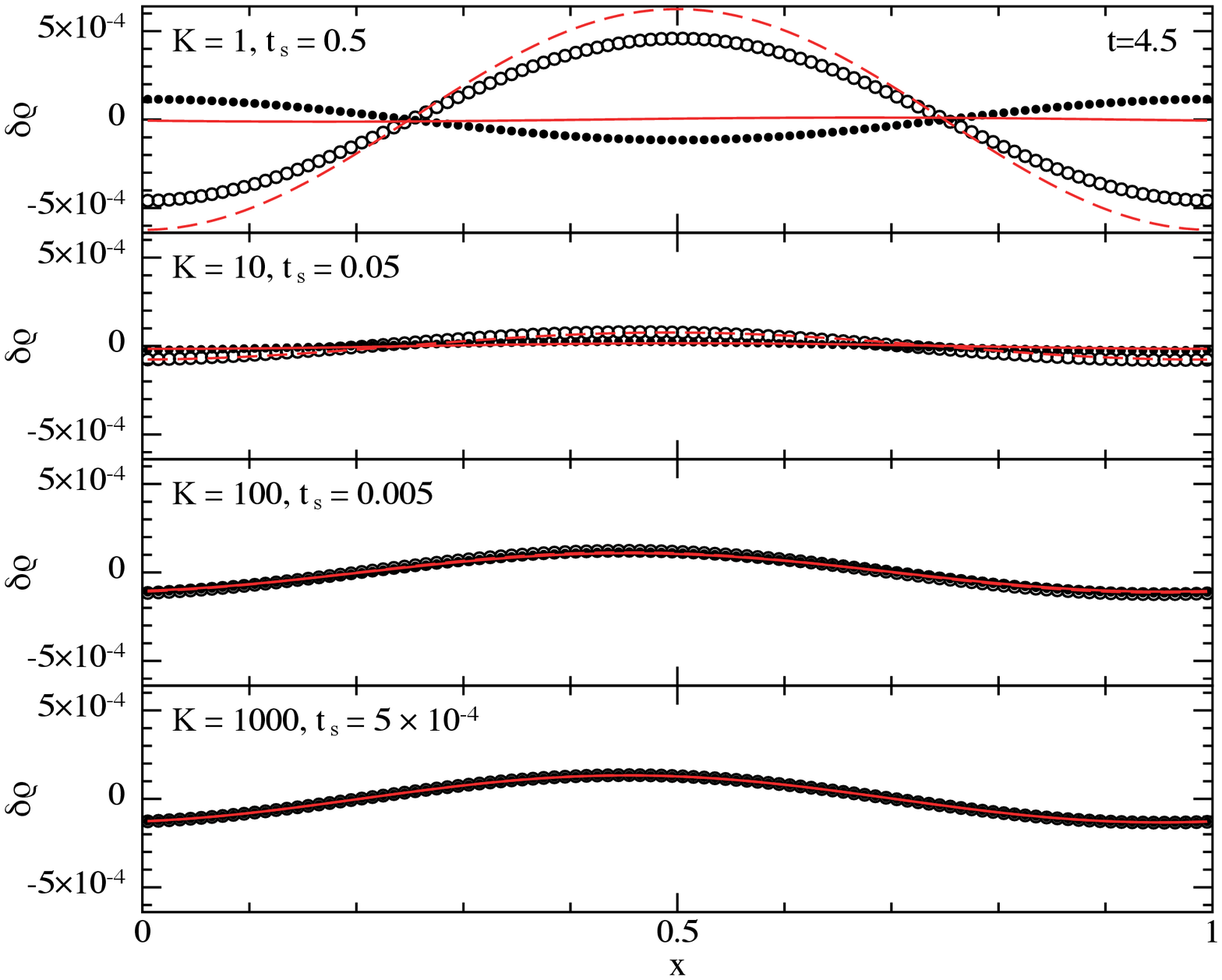}
   \caption{As in Fig.~\ref{fig:dustywave-vels} but showing the density perturbation. The solution in this case may be compared to the red solid (gas) and dashed (dust) lines, showing a high resolution non-linear solution computed using the general one fluid algorithm from \citetalias{laibeprice14a}. The solution is captured with increasing accuracy as the drag becomes stronger, with an $L_{2}$ error of $6\%$ for $K=100$ and 0.6\% for $K=1000$, but as expected becomes inaccurate in the regime where the approximation breaks down ($\ts \gtrsim 0.012$).}
   \label{fig:dustywave-dens}
\end{figure*}

There is a fundamental inconsistency in the \textsc{dustywave} initial conditions when using the terminal velocity approximation because the setup of the problem and hence the analytic solution assumes that $\Delta \bm{v}_{0} = 0$. By definition in the terminal velocity approximation we have $\Delta \bm{v} \equiv \ts \Delta \bm{f}$ which is non-zero. Hence the solution even at $t=0$ is not identical to the full one fluid case. However, these differences become smaller at large drag and at later times.

\begin{figure*} 
   \centering
   \includegraphics[width=\columnwidth]{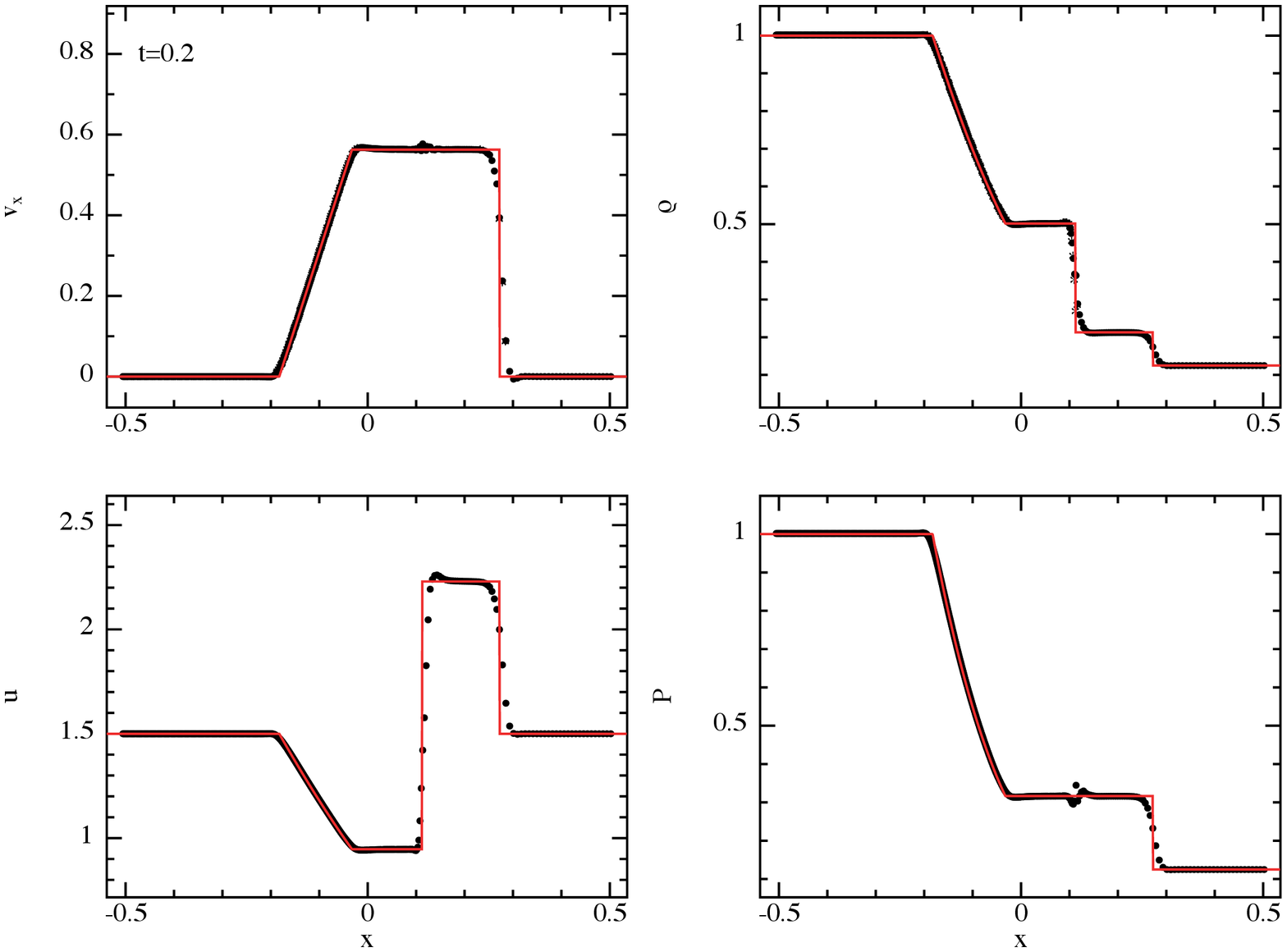}
   \includegraphics[width=\columnwidth]{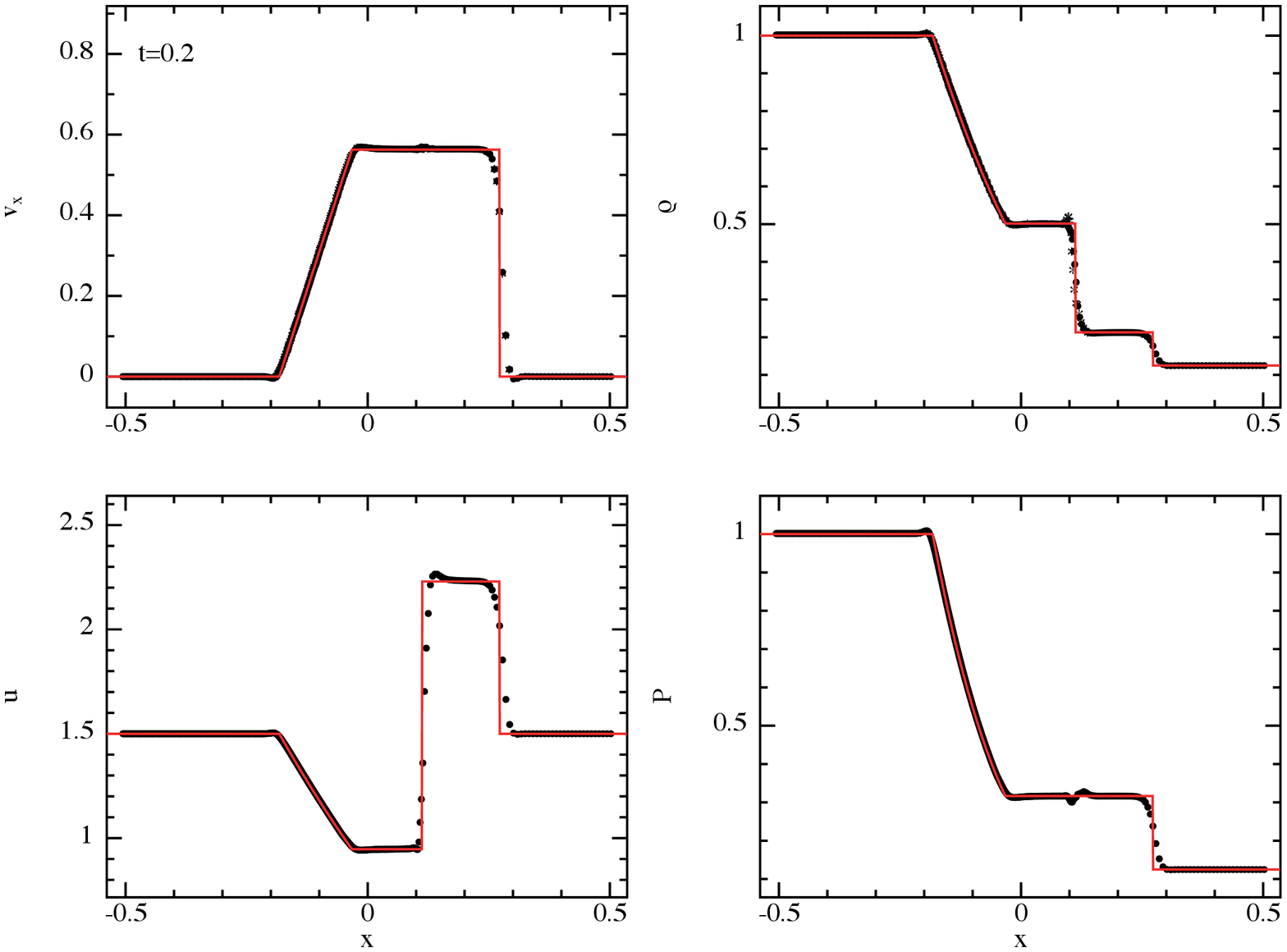}
   \caption{Results of the \textsc{dustyshock} test with a large drag coefficient, K=1000, comparing the use of two first derivatives to compute the dust diffusion (left) with the direct second derivative discretisation of the diffusion term (right). In both cases the numerical solutions agree with the analytic solution valid in the limit of infinite drag (solid red line), although the pressure is smoother across the contact discontinuity when the diffusion term is computed directly. The advantage of the present scheme compared to the full one fluid approach \citepalias{laibeprice14a} is that we have used explicit timestepping. We also avoid the punitive $h < \ts c_{\rm s}$ resolution requirement associated with the two fluid formulation \citepalias{laibeprice12}. 569 SPH particles were used.}
\label{fig:dustyshock}
\end{figure*}

 The numerical solution is shown after 4.5 wave periods in Figure~\ref{fig:dustywave-vels}, showing the gas and dust velocities (filled and open circles, respectively). As in \citetalias{laibeprice14a} we have reconstructed the gas and dust velocities on each particle from the barycentric variables, i.e. $\bm{v}_{\rm g} \equiv \bm{v} - \epsilon \Delta \bm{v}$ and $\bm{v}_{\rm d} \equiv \bm{v} + (1 - \epsilon) \Delta \bm{v}$. The left Figure shows the results using the two first derivatives approach (Sec.~\ref{sec:firstderivs}) while the right Figure shows the results using the direct second derivatives version (Sec.~\ref{sec:2ndderiv}) in each case compared to the linear analytic solution from \citetalias{laibeprice11}. There is no distinguishable difference between the two approaches. The solution in the regime where the terminal velocity approximation is valid ($K \gtrsim 42$; lower two panels in each Figure, corresponding to $\ts > \Delta t_{\rm Cour} = 0.01$) is within a few percent of the analytic solution. There is a conspicuous phase error at lower drag ($K = 10$ and $K=1$; first and second row), in part caused by the inconsistency in the initial conditions, which becomes worse as $\ts$ becomes larger, and in part because this is where the terminal velocity approximation breaks down. Nevertheless the general behaviour in terms of the damping of the wave at intermediate drag is captured despite the inapplicability of the approximation in this regime. The behaviour at even lower drag ($K < 1$; not shown) is incorrect; here the wave remains damped when using the terminal velocity approximation whereas the damping should decrease as the coupling tends to zero. Hence the full one fluid approximation should be used in this regime \citepalias[e.g.][]{laibeprice14a}, as we argued in Sec.~\ref{sec:timestepping}.
 
 Figure~\ref{fig:dustywave-dens} shows the solution for the density perturbation. Importantly, the analytic solution for density in the \textsc{dustywave} problem quickly becomes nonlinear, particularly when the drag is weak. This can be seen by considering the limit of no drag: Assuming the dust is not submitted to any external force we have
\begin{equation}
v(t) = v(t = 0) = v_{0} \sin \left(k x_{0} \right) ,
\end{equation}
implying
\begin{equation}
x(t) = x_{0} - v_{0}\sin \left(k x_{0} \right) t .
\end{equation}
Hence, from mass conservation, the dust density is given by
\begin{equation}
\rhod (t) = \frac{\rho_{\mathrm{d}0}\left(x_{0} \right)}{\displaystyle \left| \left(\frac{\partial x}{\partial x_{0}} \right)_{t} \right|} = \frac{\rho_{\mathrm{d}0}\left(x_{0} \right)}{\left| 1 - v_{0}k \cos\left(k x_{0} \right) t \right|}.
\end{equation}
This result is physically consistent with the initial velocity profile: grains are depleted at $x = \pm \pi$, pile up at $x = 0$ and maintain a constant density at $x = \pm \pi / 2$ (zero net flux of particles). In particular, density fluctuations become of the order of the background on a typical time $\left( v_{0}k\right)^{-1}$ and the analytic solution of the \textsc{dustywave} problem from \citetalias{laibeprice11} cannot be applied anymore. It should be noted however that the velocities remain small and still agree with the solution of the linear problem. Hence, we have computed the reference solution in Fig.~\ref{fig:dustywave-dens} using a high-resolution (5000 particle) simulation with our fully general one fluid algorithm \citepalias{laibeprice14a}; whereas in Fig.~\ref{fig:dustywave-vels} we used the linear solution from \citetalias{laibeprice11} (both methods produce indistinguishable results for the velocity field).

Since there is no inconsistency in the density in the initial conditions, the solution using the diffusion approximation is more accurate for the densities than for the velocities ($L_{2}$ error of $0.06$ at $K=100$ and $0.006$ at $K=1000$), though still becomes inaccurate ($L_{2}$ error $\gtrsim 0.5$) for $K \leq 10$. As with the velocities, there is no difference between the two implementations (compare left and right panels in Figure~\ref{fig:dustywave-dens}), indicating that any inaccuracies are due to the physical approximation rather than the numerical scheme itself.

\subsection{\sc Dustyshock}
 The \textsc{dustyshock} problem at strong drag was one of the most difficult problems to solve using a two fluid approach due to the resolution requirement $h \lesssim c_{\rm s}\ts$ that leads to overdamping of the solution if not satisfied \citepalias{laibeprice12}. We have already shown in \citetalias{laibeprice14a} that this spatial resolution requirement is unnecessary when using a general one fluid formulation, although the drag still imposes a prohibitive timestep constraint, meaning that an implicit timestepping scheme (albeit a fairly simple one) is still necessary. Figure~\ref{fig:dustyshock} show that with our present method we can capture the high-drag \textsc{dustyshock} solution using explicit timestepping without any timestep constraint other than the usual Courant condition, and without any particular spatial resolution requirements.

 We set up the problem as usual, following the standard \citet{sod78} shock tube with conditions in the gas for $x \leq 0$ given by ($\rho_{\rm g}, v_{\rm g}, P) = (1.0, 0.0, 1.0)$ and for $x > 0$ given by ($\rho_{\rm g}, v_{\rm g}, P) = (0.125, 0.0, 0.125)$. We assume a constant dust fraction in the initial conditions ($\epsilon_{0} = 1$), using 569 particles (corresponding to a particle spacing of $\Delta x = 0.001$ for $x \leq 0$, an adiabatic equation of state with $\gamma = 5/3$ and a drag coefficient $K=1000$. In this regime the solution corresponds to the usual hydrodynamic solution with a modified sound speed (red lines in Figure~\ref{fig:dustyshock}). In this respect we are testing only the ability of the algorithm to recover the zeroth order effect of a heavy fluid \citepalias[see][]{laibeprice14}, which from the results in Figure~\ref{fig:dustyshock} can be seen to be true.

  As previously there is very little difference between the two implementations (comparing left and right panels) except that the direct second derivatives approach (c.f. Section~\ref{sec:2ndderiv}) produces slightly less noise in the pressure profile across the contact discontinuity. While this can be important for some problems \citep[see e.g.][]{price08}, such a minor difference is not enough to prefer this discretisation over the two first derivatives approach. However, given that the direct second derivatives algorithm is also significantly faster it may be preferred on this basis.
  \begin{figure*}
   \centering
   \includegraphics[width=\columnwidth]{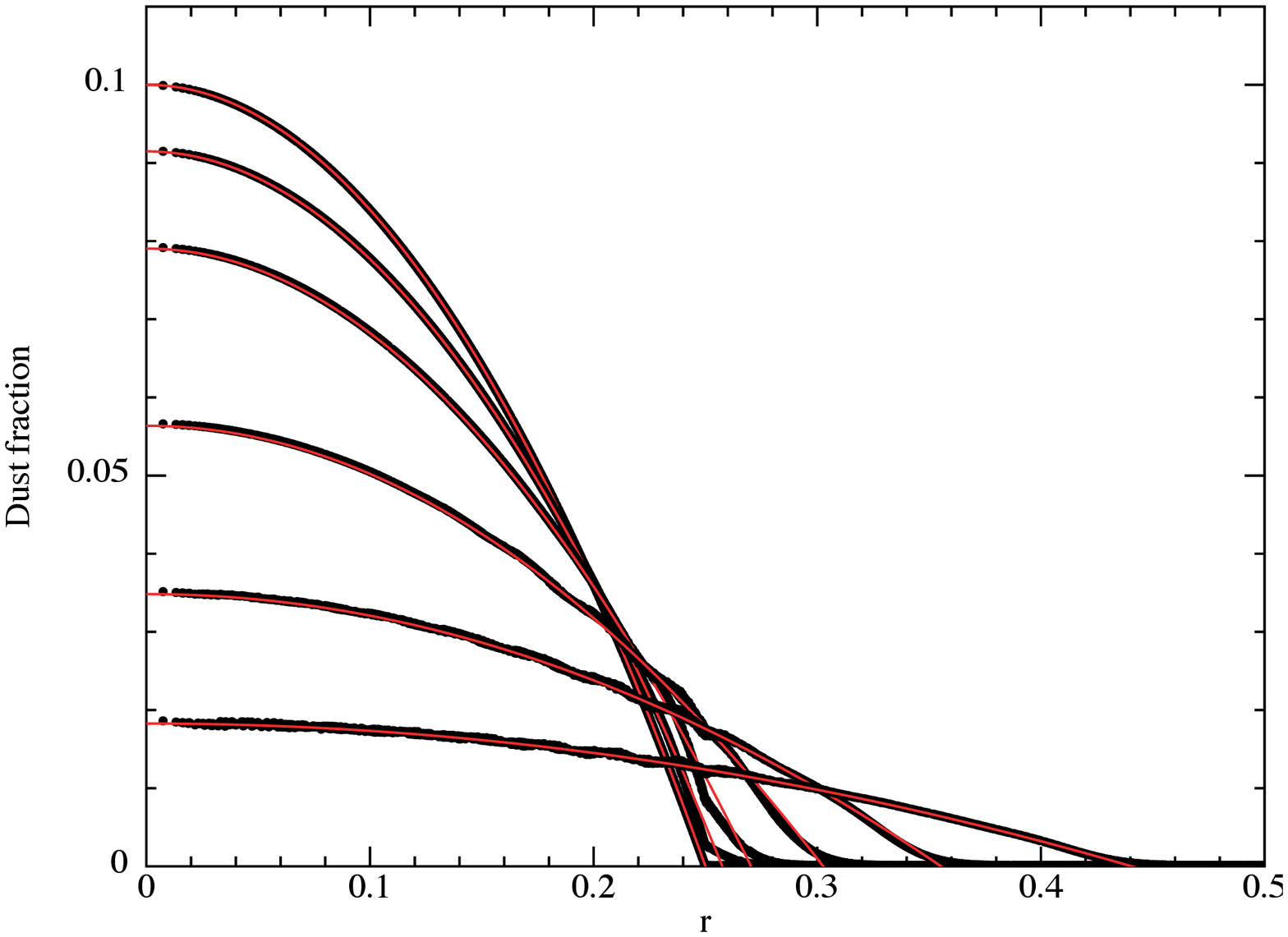}
   \includegraphics[width=\columnwidth]{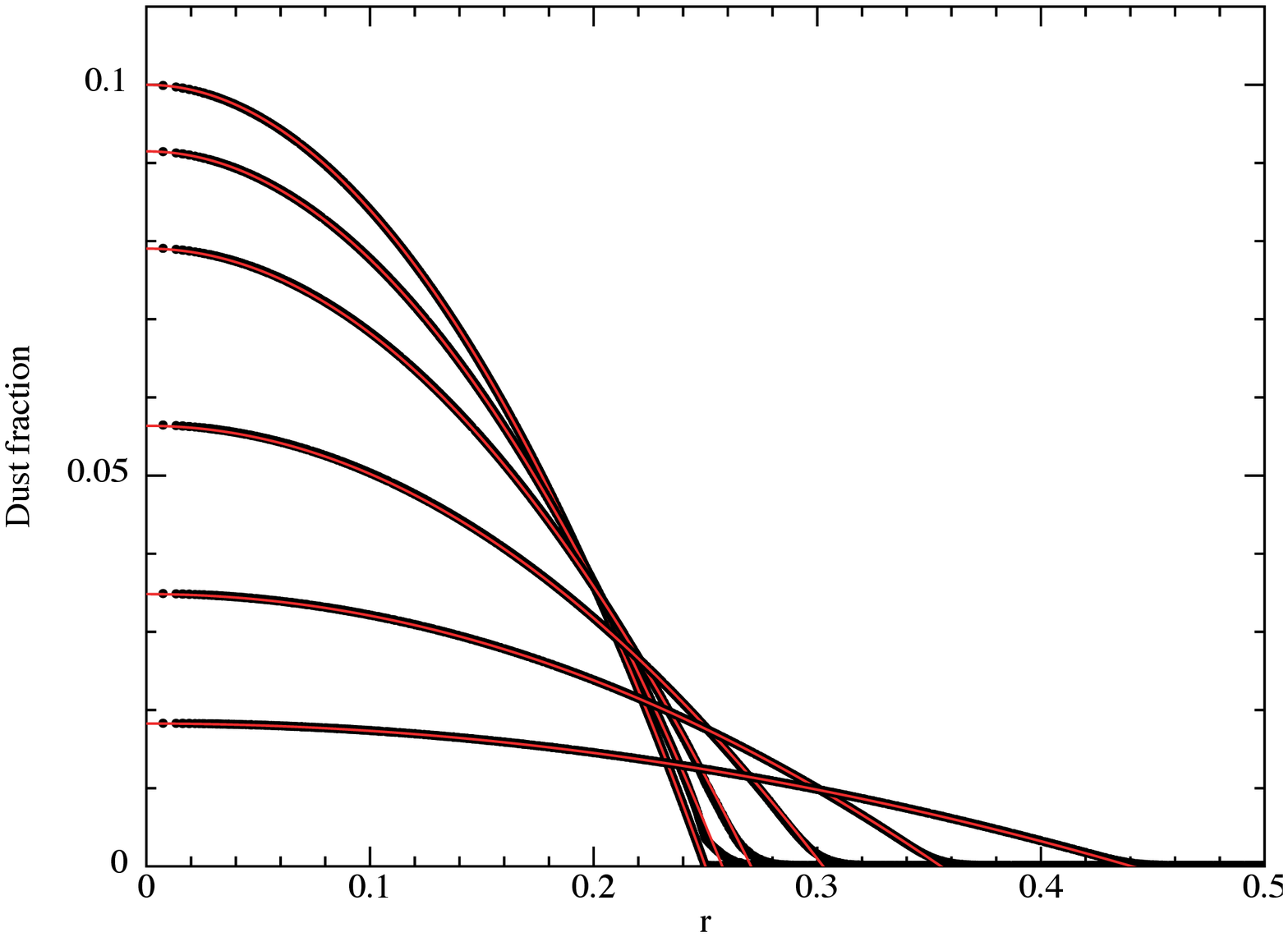}
   \caption{Dust fraction as a function of spherical radius in the 3D dust diffusion test at $t=0.0$, 0.1, 0.3, 1, 3 and 10 (top to bottom) from simulations using $50 \times 58 \times 60$ particles. The numerical solution, projecting all particles in $r$, is given by the black dots and may be compared to the analytic solution given by the red lines. The left panel shows the solution with two first derivatives, while the right panel uses the direct second derivative.}
\label{fig:diffuse}
\end{figure*}

\subsection{\sc Dustydiffusion}
\label{sec:dustydiffusion}
 Based on Equation~\ref{eq:diffuse}, we present a new test for dust-gas mixtures with a simple analytic solution. This consists of the steady diffusion of an overconcentration of dust. To set up the problem we consider a uniform density box with $\rho = \rho_{0} = 1$ and an isothermal equation of state $P = c_{\rm s}^{2} \rho_{\rm g}$ with $c_{\rm s} = 1$. In this case the dust diffusion can be described by Equation~\ref{eq:diffuse}. For the diffusion parameter we assume that the stopping time is a constant (this is equivalent to assuming an Epstein-like drag where $t_{\rm s} = \rho_{\rm grain} s_{\rm grain}/(\rho c_{\rm s})$ is constant). 

\subsubsection{Analytic solution}
The exact solution can be obtained by solving the equation
\begin{equation}
\frac{\mathrm{d}\epsd}{\mathrm{d}t} = \nabla\cdot\left(\epsilon \tilde{\eta} \nabla \epsilon  \right),
\label{eq:epsdiff}
\end{equation}
where $\tilde{\eta} \equiv  t_{\rm s} c_{\rm s}^{2}$ is a constant. We solve this by assuming spherical symmetry, i.e.
\begin{equation}
\frac{\mathrm{d}\epsd}{\mathrm{d}t} = \frac{\tilde{\eta}}{r^{2}} \frac{\rm d}{{\rm d}r} \left( r^{2}\epsilon \frac{{\rm d} \epsilon}{{\rm d} r} \right),
\label{eq:epsdiffr}
\end{equation}
for which there are several known analytic solutions, including the general time-dependent solution
\begin{equation}
\epsilon(r,t) = A \left\vert 10 \tilde{\eta} t+B \right\vert^{-\frac{3}{5}} - \frac{r^2}{ 10 \tilde{\eta} t+B} , \label{eq:sol3}\\ 
\end{equation}
where $A$ and $B$ are arbitrary constants. We use this solution to verify our numerical scheme by solving only the diffusion equation via either Equations~\ref{eq:spheps} and \ref{eq:deltafg} or Equation~\ref{eq:depsdtsph}, with the particle positions fixed (for this problem only)\footnote{We attempted to construct an equilibrium situation involving all of Equations~\ref{eq:ctyalt}--\ref{eq:epsalt}, for example a hydrostatic equilibrium in a fixed potential. However, it is difficult to construct an equilibrium where the dust simply diffuses according to (\ref{eq:epsdiff}) because the change to $\epsilon$ causes a change to the pressure gradient and hence causes an acceleration to the barycentre also.}.
 
 \begin{figure} 
   \centering
   \includegraphics[width=\columnwidth]{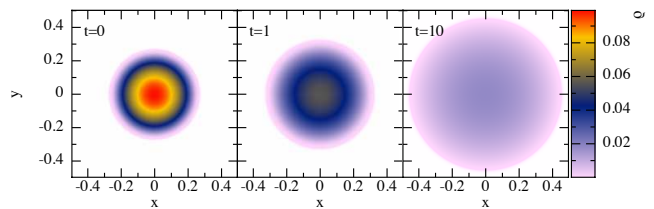}
   \caption{Cross section of the dust density in the $z=0$ plane in the 3D dust diffusion test at $t=0$, 1 and 10 (left to right).}
\label{fig:diffuse-dens}
\end{figure}

\subsubsection{Results}
  We set up the problem in 3D with $50\times 58 \times 60$ particles set on a uniform close-packed lattice in the domain $x,y,z \in [-0.5, 0.5]$. The positions of the y and z boundaries are adjusted slightly to ensure periodicity of the lattice across the boundary (the particle spacing in $x$, $y$ and $z$ is $\Delta p$, $\sqrt{3}\Delta p/2$ and $\sqrt{6}\Delta p/3$ respectively, where $\Delta p = 0.02$). We use an isothermal equation of state, setting $c_{\rm s} = 1$ and $t_{\rm s} = 0.1$ such that $\tilde{\eta} = 0.1$, and set the initial dust fraction using
 \begin{equation}
 \epsilon(r,0) = \epsilon_{0}\left[1 - \left(\frac{r}{r_{c}}\right)^{2}\right],
 \end{equation}
consistent with Equation~\ref{eq:sol3} with $B \equiv \epsilon_{0}/r_{\rm c}^{2}$ and $A \equiv \epsilon_{0} B^{\frac35}$. We set $\epsilon_{0} = 0.1$ and $r_{\rm c} = 0.25$.

  Figure~\ref{fig:diffuse} compares the numerical solution to the analytic solution, while Figure~\ref{fig:diffuse-dens} illustrates the general behaviour of the solution. The solution with Equation~\ref{eq:depsdtsph} (right panel of Fig.~\ref{fig:diffuse}) is excellent ($L_{\rm 2}$ error $\lesssim 5 \times 10^{-4}$ for $r < 0.2$), apart from the physical deviation from the self-similar solution due to the transition to constant rather than negative $\epsilon$ at the outer radius. The solution with using Equations~\ref{eq:spheps} and \ref{eq:deltafg} (left panel) is also good, but shows some low amplitude oscillations that develop from the propagation of the `kink' in the initial epsilon profile. These oscillations are worse at lower resolution (they can be smoothed out by adding some artificial dissipation in $\epsilon$ but the solution is still not as good as using Equation~\ref{eq:depsdtsph}). 
  
\begin{figure}
   \centering
   \includegraphics[width=0.9\columnwidth]{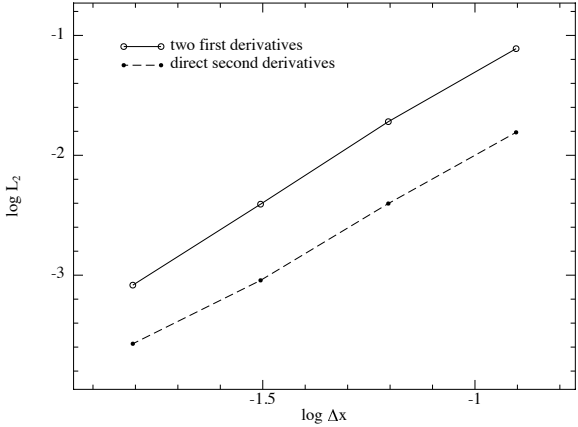}
   \caption{Convergence in the dust diffusion problem, showing $L_{2}$ error for the solution within $r < 0.2$ as a function of the particle spacing. While both methods show second order convergence, the direct second-derivatives solution is more accurate because of oscillations in the two first derivatives approach propagating from the `kink' in the initial $\epsilon$ profile seen in Figure~\ref{fig:diffuse}.}
\label{fig:conv}
\end{figure}

  Figure~\ref{fig:conv} quantifies these results with a convergence study using $8^{3}$, $16^{3}$, $32^{3}$ or $64^{3}$ particles arranged on a cubic lattice. We show the $L_{2}$ error computed by \textsc{splash} \citep{price07} from particles with $r < 0.2$. While in both cases the convergence is second order $\propto (\delta x)^{2}$, it can be seen that the direct second derivatives approach gives results more accurate by a factor of $5$ at any given resolution. Our results with both schemes when employing $s$ instead of $\epsilon$ (Appendix~\ref{sec:s}) are worse by a factor of $\sim2$, again with a similar preference for the direct second derivatives approach. Thus while it is clear that all of our proposed numerical schemes correctly discretise the diffusion equation, we find the discretisation using Equation~\ref{eq:depsdtsph} to be more accurate for this problem.

%

\subsection{Dust settling in a protoplanetary disc}
\label{sec:settle}

Our final test is drawn from our intended application, namely the dynamics of small grains in protoplanetary discs. 
\subsubsection{Setup}
We simplify the problem by considering only the vertical settling of grains in the r-z plane. That is, we set up particles in a two dimensional cartesian box with an acceleration in the `vertical' (z) direction given by
\begin{equation}
a_{z} = -z \frac{GM}{\left(R_{0}^{2} + z^{2}\right)^{\frac32}},
\label{eq:acc_y}
\end{equation}
where we assume code units such that $GM=1$ and set $R_{0}=5$ as a constant. The boundary conditions are periodic in the horizontal (x) direction and free in the vertical direction. We use an isothermal equation of state $P=c_{\rm s}^{2} \rho$ where the sound speed $c_{\rm s}$ is set such that the aspect ratio $H/R_{0} \equiv c_{\rm s}^{2}/(\Omega_{0} R_{0}) = 0.05$, where $\Omega_{0} \equiv \sqrt{GM/R_{0}^{3}}$. The orbital time is therefore $t_{\rm orb} \equiv 2\pi/\Omega_{0} \approx 70$ in code units. We set particles of equal mass initially on a uniform hexagonal lattice in the domain $x \in [-0.25, 0.25]$ and $z \in [-3H,3H]$. We specify the particle separation in the $x$ direction to be either 16, 32, or 64, resulting in $16 \times 56 = 856$ particles at the lowest resolution, $32 \times 111 = 3552$ particles at medium resolution and $64 \times 222 = 14208$ at the highest resolution.

 We then stretch the particle distribution to match the equilibrium density profile using the method described in \citet{price04} where the $z$ position of each particle is determined by solving the root finding problem
\begin{equation}
f(z) = \frac{M(z)}{M(z_{\max})} - \frac{(z_{0} - z_{\min})}{z_{\max} - z_{\min}} = 0,
\end{equation}
where $M(z) \equiv \int_{z_{\min}}^{z} \rho(z') {\rm d}z'$, $z_{0}$ is the initial position of the particle and we set
\begin{equation}
\rho(z) = \rho_{0} \exp[-z^{2}/(2H^{2})]. \label{eq:rhoz}
\end{equation}
We set the mass of each particle equal to $M(z_{\rm max})$ divided by the number of particles in the domain, consistent with the desired density profile. Equation~\ref{eq:rhoz} is a slight approximation (fourth order in $z/H$; e.g. \citealt{LGM12}) but this is unimportant since we relax the particles into a hydrostatic equilibrium anyway, as described below.

 We set up the simulation initially with only gas and run the calculation to $t=1000$ in code units (i.e. $\sim$14 orbits) with both artificial viscosity and an artificial damping term of the form 
\begin{equation}
\left(\frac{{\rm d}\bm{v}}{{\rm d}t} \right)_{\rm damp} = -f_{\rm damp}\bm{v},
\end{equation}
where $f_{\rm damp} = 0.03$ in order to allow the distribution to relax to equilibrium. We then add dust to the simulation, assuming a dust-to-gas ratio of $\rho_{\rm d}/\rho_{\rm g} = 0.01$ by setting the dust fraction using
\begin{equation}
\epsilon \equiv \frac{\rho_{\rm d}}{\rho} = \frac{\rho_{\rm d}/\rho_{\rm g}}{( 1 + \rho_{\rm d}/\rho_{\rm g})}.
\end{equation}
We then evolve the simulation for a further 50--100 orbits.

\begin{figure*}
   \centering
   \includegraphics[width=\textwidth]{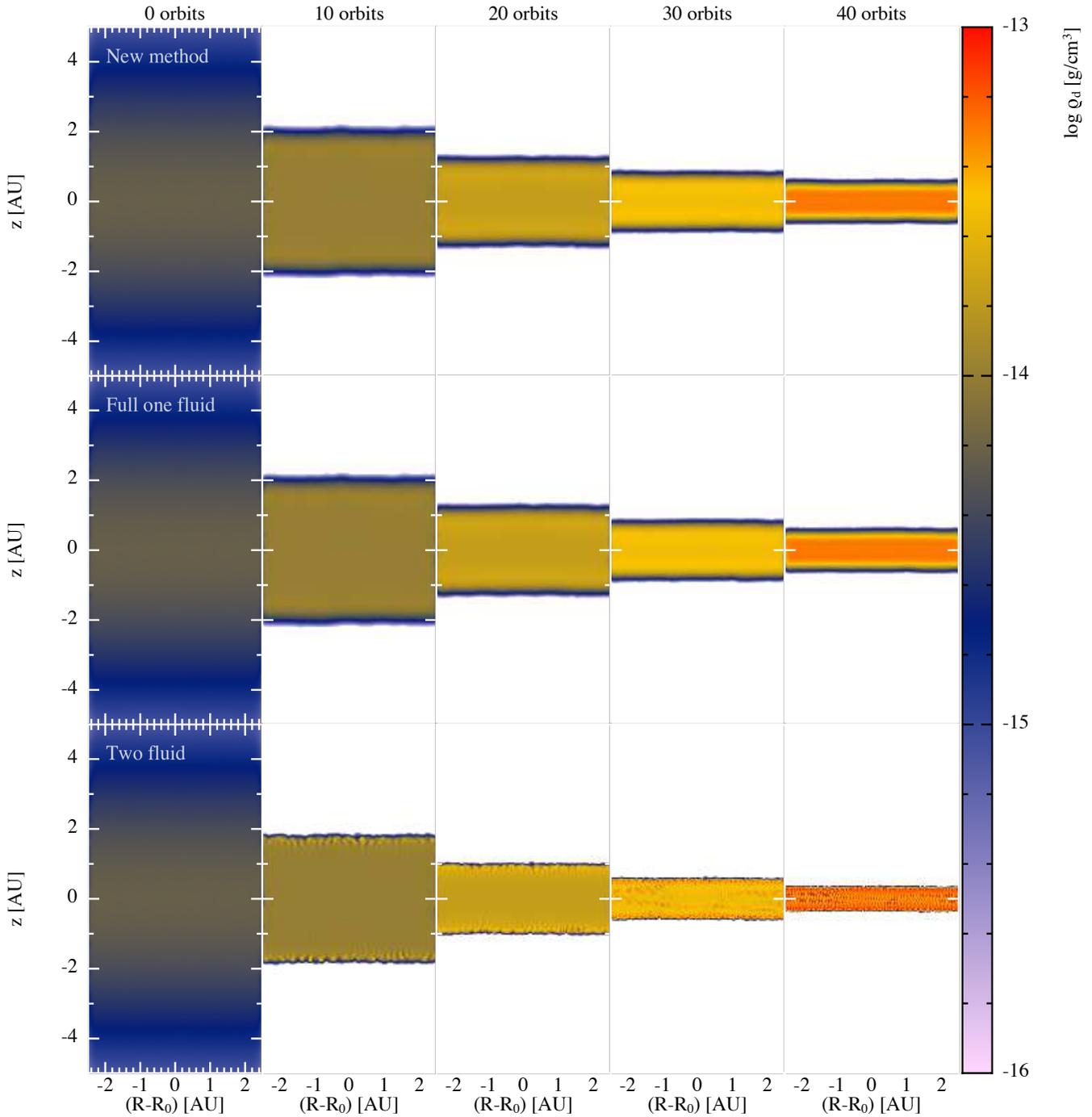}
   \caption{Settling of mm dust grains in a 2D (r-z) vertical section of a protoplanetary disc at $R_{0}$ = 50AU (assuming $H/R = 0.05$; so $H_{0} = 2.5$AU) using $32 \times 111$ mixture particles. The plot shows dust density as a function of time. The top row shows the results using our new dust diffusion method. The solution may be compared to that obtained with the full one fluid formulation from \citet{laibeprice14a} (middle row) and with the two fluid formulation \citet{laibeprice12,laibeprice12a} (bottom row; uses $32 \times 111$ particles in both gas and dust). Our new method requires half the number of particles compared to the two fluid approach and is $50$ times faster.}
\label{fig:settle}
\end{figure*}

  To give the problem physical meaning we consider a distance unit of $10$AU (such that $R_{0} = 5$ corresponds to 50AU), a mass unit of $1 M_{\rm \odot}$ and the time unit set such that $G=1$ in code units. This implies an orbital time of $2\pi/\Omega_{0} = 353$ years. A midplane density $\rho_{0}$ of $10^{-3}$ in code units then corresponds to $\approx 6 \times 10^{-13}$~g/cm$^{3}$, giving a disc surface density $\Sigma \approx 55$~g/cm$^{2}$. We adopt a linear Epstein drag prescription, defining the stopping time according to Eq.~\ref{eq:tseps}. We set the intrinsic grain density $\rho_{\rm grain} = 3$~g/cm$^{3}$. The midplane stopping time at $R_{0}$ is given by
\begin{equation}
t_{\rm s} \Omega_{0} = 1.35 \times 10^{-3} \left(\frac{s_{\rm grain}}{1 {\rm mm}}\right) \left(\frac{\rho_{\rm grain}}{3 {\rm g}/{\rm cm}^{3}}\right) \left(\frac{\rho}{\rho_{0}} \right)^{-1}.
\end{equation}

\subsubsection{Settling of millimetre grains}
We first perform a series of tests with a grain size of $s_{\rm grain} = 1$~mm, chosen as a balance between the regime where the diffusion method is applicable and where it is still possible to obtain a solution in a reasonable time with the two fluid method. For the setup above this is at the limit of where the diffusion method is applicable, and indeed we found that Eq.~\ref{eq:dtd} controlled the timestep, indicating that the time dependence of the differential velocity has started to become important. In \citetalias{laibeprice12} we showed that it was necessary to satisfy $h \lesssim c_{\rm s} t_{\rm s}$ to avoid overestimating the drag. For 1mm grains our two fluid calculations violate this criterion by a factor of $\sim 9$, 4.5 and 2.25 at the midplane at low, medium and high resolution, respectively, but do not appear to show overdamping. \citet{lorenbate14} found that the resolution problem is not as severe when the dust-to-gas ratio is low, suggesting that $h \leq \epsilon c_{\rm s} t_{\rm s}$ is a more precise resolution criterion.

 \begin{figure}
   \centering
   \includegraphics[width=\columnwidth]{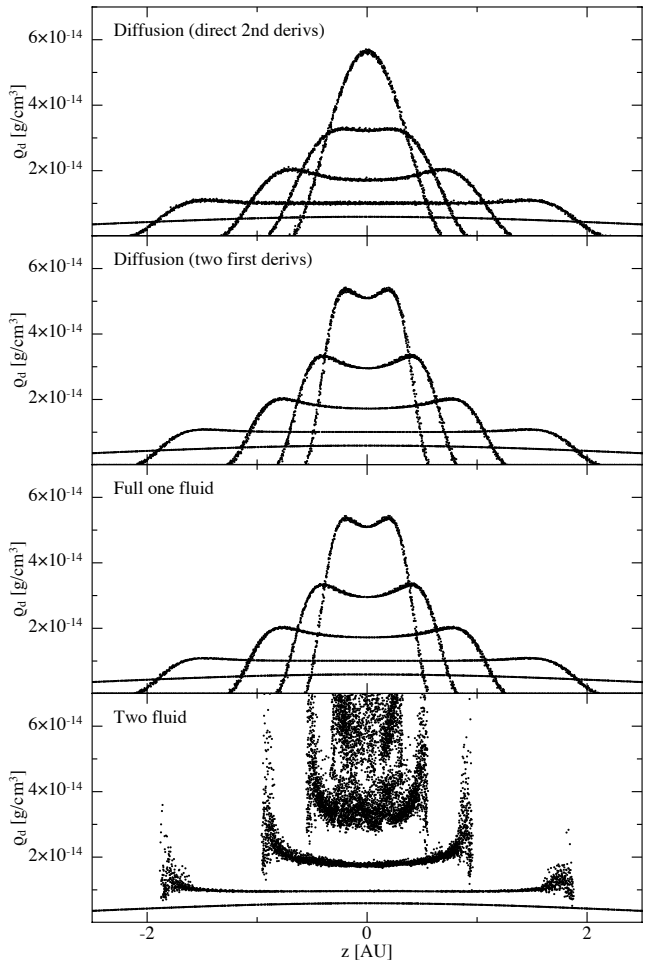}
   \caption{As in Figure~\ref{fig:settle} but showing the projection of dust density on all particles in 2D as a function of $z$, plotted at $t=0$,10,20,30 and 40 orbits for the different methods. The direct second derivatives formulation (top) gives a slightly more smoothed result compared to the two first derivatives method (second row), the latter of which is indistiguishable from the solution with the full one fluid method (third row). The one fluid methods are less well resolved in the dust density for this problem but are give a significantly less noisy solution than obtained with the two fluid method (bottom). The diffusion method is 50 times faster at this resolution and grain size, with increasing performance gains for smaller grains.}
\label{fig:rhoz}
\end{figure}

 \begin{figure}
   \centering
   \includegraphics[width=\columnwidth]{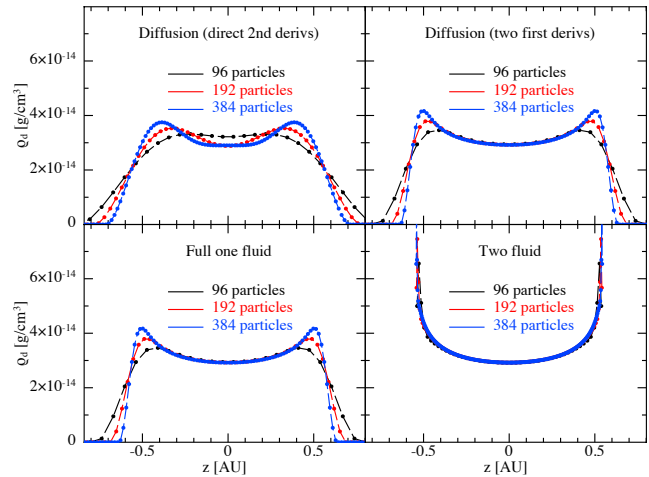}
   \caption{Resolution study in 1D version of dust settling problem (Figs.~\ref{fig:settle} \& \ref{fig:rhoz}), showing solution after 30 orbits using the diffusion approximation with direct second derivatives (top left), two first derivatives (top right), the one fluid method from \citetalias{laibeprice14a} (bottom left) and the two fluid method (bottom right). The particle number refers to the total number of particles in the domain in each case. In the one fluid methods the resolution follows the \emph{total} mass rather than the dust mass, so the dust density is comparatively less well resolved.}
\label{fig:resstudy}
\end{figure}

Figure~\ref{fig:settle} shows the dust density in the medium resolution calculations at intervals of ten orbital periods using three different methods. The top row shows the results with our new method employing the two first derivatives approach (Section~\ref{sec:firstderivs}). The results with direct second derivatives (Equation~\ref{eq:depsdtsph}) are similar but slightly less well resolved (see Fig.~\ref{fig:rhoz}). The second row of Fig.~\ref{fig:settle} shows the solution obtained with the general one fluid method from \citetalias{laibeprice14a}. The main difference is that the differential velocity $\Delta \bm{v}$ is explicitly evolved in that formulation, and so there is a timestep constraint from the stopping time which makes the simulation run $\sim 25$ times slower for this grain size when computed with only explicit timestepping with the constraint $\Delta t < t_{\rm s}$. With the diffusion approximation the timestep constraint is \emph{inversely} proportional to the stopping time although quadratically proportional to resolution (Equation~\ref{eq:dtd}). The third row shows the solution obtained with the two fluid algorithm \citepalias{laibeprice12}. In this case instead of setting the dust fraction we added a separate set of dust particles copied from the gas particles but with 1\% of the mass. This approach therefore required twice the number of particles compared to our diffusion algorithm and the timestep is also constrained by the stopping time. Computing this solution required approximately 50 times more cpu time than the diffusion method.
 
All three methods produce dust settling on a comparable timescale, with only minor differences in the numerical solutions. A more detailed comparison is given in Figure~\ref{fig:rhoz}, showing the dust density on all particles as a function of $z$ at the same times as those shown in Figure~\ref{fig:settle}. The main noticeable difference is that the two fluid solution contains more noise in the particle distribution. This is because the dust is modelled as a separate set of particles that feel no mutual repulsion, compared to the one fluid case where the dust distribution benefits from the regular arrangement of the mixture particles. The approach with direct second derivatives (top row) produces a slightly over-smoothed solution compared to the the two first derivatives approach (second row) --- with the latter giving results that are indistinguishable from the full one fluid method (third row), showing that the diffusion approximation is indeed accurate in this regime.

 A major difference between the one fluid methods and the two fluid method is that resolution is tied to the \emph{total} mass rather than the dust mass. This is evident in Figure~\ref{fig:rhoz} where the two fluid method (bottom row) can be seen to better capture the ``wings'' in the dust density at high latitudes. We quantify this further in Figure~\ref{fig:resstudy} with a resolution study of the same problem performed in 1D to avoid the particle noise in the two fluid approach. Settling means that after some time the dust covers a much smaller region of the domain than the gas, so the one fluid formulations under-resolve the dust compared to the two fluid method, since resolution is tied to the total mass, most of which remains at high latitudes. By contrast in the two fluid approach resolution is tied to the dust mass and so is naturally placed towards regions of high dust density. This is both an advantage and a disadvantage to both types of approaches, it depends whether it is desirable to resolve the total mass or the dust mass. This point is discussed in \citetalias{laibeprice14a} mainly as an advantage to the one fluid method since it avoids the possibility of dust particles becoming `trapped' below the resolution of the gas.

 \begin{figure*}
   \centering
   \includegraphics[width=\textwidth]{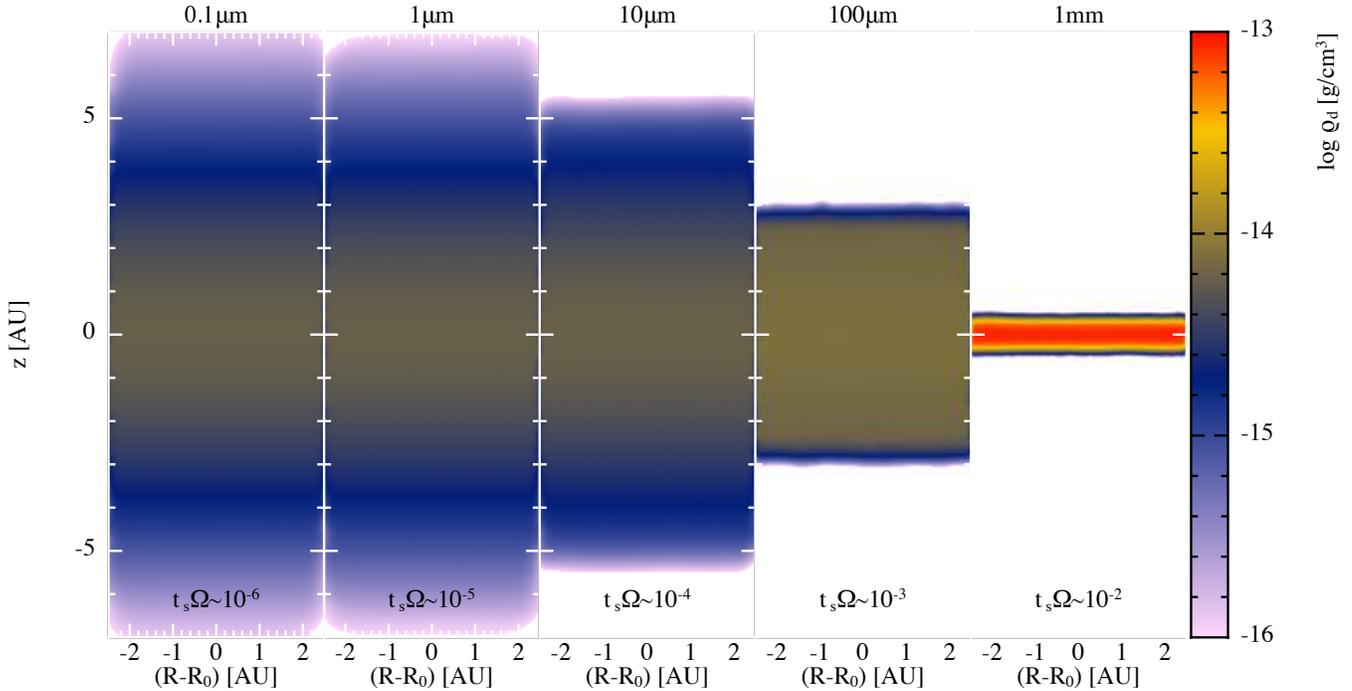}
   \caption{As in Figure~\ref{fig:settle} but comparing settling of grains of different sizes at in a vertical section of a protoplanetary disc at 50 AU. Each panel shows the dust density after 50 orbits computed with our new method (similar results are found with both implementations described in this paper). The decrease in settling time with increasing grain size (left to right) is clearly evident. Simulations in this regime are prohibitively slow with the two fluid approach. We used explicit timestepping in all cases, with the simulations for grain sizes of $<$ 1mm constrained only by the Courant condition.}
\label{fig:grains}
\end{figure*}

 In obtaining our results with the diffusion approximation, we found a few caveats to the numerical algorithm we derived in Section~\ref{sec:numerics}. First, we found it necessary with both to use the $M_{6}$ kernel for this problem to obtain a smooth and accurate solution (this kernel extends to $3h$ instead of $2h$ and so better approximates the Gaussian). To avoid particle pairing occurring at high latitudes with the quintic we set the ratio of smoothing length to particle spacing to 1.0 instead of 1.2, equivalent to using a mean neighbour number of 28.3 in 2D (see \citealt{price12}). The second caveat was that the dust fraction becomes negative around the edge of the collapsing dust layer in both of our implementations (and also with the full one fluid method). This arises because of the exact conservation of the dust mass in the algorithm, which causes a slight overshoot at the discontinuity. In order to smoothly handle this we derived an alternative approach (Appendix~\ref{sec:s}) which guarantees a positive dust fraction, but we found it to give less accurate results than the method employing $\epsilon$ (c.f. Sec.~\ref{sec:dustydiffusion}). Instead, we found that the most effective way of solving this was to simply set the dust fraction to zero on particles where it had become negative. This slightly violates the exact conservation of the dust mass, but the error is small ($\sim 10^{-5}$ in $\epsilon$ with the quintic) and it is a small price to pay for stability of the algorithm.
 
\subsubsection{Settling with different grain sizes}
 Finally, we demonstrate the ability of the diffusion method to simulate small grains in a protoplanetary disc by performing a series of calculations varying the grain size from $0.1{\mu}$m to $1$mm. Grain sizes below 1mm are difficult to simulate at all with the two fluid technique because of the punitive spatial and temporal resolution requirements \citepalias{laibeprice12,laibeprice12a}. With the general one fluid method presented in \citetalias{laibeprice14a} the limitation on the spatial resolution is removed, because we are no longer modelling the separation between fluids with physically separate resolution elements, but it is still necessary to use implicit timestepping. Yet these grains are important in protoplanetary discs as they control much of the thermal radiation.

 Figure~\ref{fig:grains} shows the results of a series of medium resolution ($32 \times 111$) calculations of dust settling for different grain sizes, shown after 50 orbits at $R_{0}$. We used only explicit timestepping, and for grain sizes smaller than 1mm the timestep was constrained only by the Courant condition. The different settling behaviour of the different grain populations in discs is clearly evident, with the micron and sub-micron grains remaining stuck to the gas at high latitudes, the millimetre grains settling effectively to the midplane and the 100 micron and 10 micron grains having partially settled.
 
 Being able to accurately and efficiently simulate single-size small grains in discs in this manner is the first step towards modelling an evolving grain population self-consistently.

\section{Discussion and conclusions}
 We have derived and implemented a numerical scheme for describing the dynamics of small dust grains coupled to a gas, using Smoothed Particle Hydrodynamics, in the limit where the stopping time is short compared to the computational timestep. This requires solving one additional diffusion equation as well as the usual equations of hydrodynamics slightly modified by some additional terms. We derived two implementations, one where the diffusion equation is computed using two first derivatives (Section~\ref{sec:firstderivs}) and one where direct second derivatives were employed (Section~\ref{sec:2ndderiv}). We found only minor differences between the two approaches on the test problems we tried. Given this, we recommend the direct second derivatives approach (Sec.~\ref{sec:2ndderiv}), which is both simpler and faster because it does not require an extra loop over the particles.
  
  As discussed in Sec.~\ref{sec:timestepping}, the terminal velocity approximation or, as we prefer, the ``diffusion approximation for dust'', is valid when the stopping time is less than the computational timestep. The simple way to guarantee this validity in practice is to ensure that the diffusion timestep (Eq.~\ref{eq:dtd}) is not constraining the timestep, otherwise the more general one fluid approach implemented in \citetalias{laibeprice14a} where the time dependence of $\Delta \bm{v}$ is kept should be used instead. In this sense the method we have described is complementary to both the full one fluid approach \citepalias{laibeprice14a} and the two fluid approach \citepalias{laibeprice12}. The main difference is that the other methods need implicit timesteps when the grain size is \emph{small} ($\ts < \Delta t$), whereas this method requires implicit timesteps when the grain size is \emph{large} ($\ts > \Delta t$), but this is where the approximation breaks down anyway.

Finally, we considered only one grain size at a time in this paper. We have recently generalised our one fluid formulation to describe an arbitrary number of grain populations all within a single fluid mixture \citepalias{laibeprice14b}. Our next step will be a numerical implementation of this more general formulation, including the simplification to a diffusion approximation, as well as modelling the evolution of the grain population including growth and fragmentation.

\section*{Acknowledgments}
DJP is very grateful for funding via an Australian Research Council (ARC) Future Fellowship, FT130100034, and Discovery Project grants DP1094585 and DP130102078. GL acknowledges funding from the European Research Council via FP7 ERC advanced grant project ECOGAL. We thank Mark Hutchison, Joe Monaghan and Giovanni Dipierro for useful discussions, and the anonymous referee for comments that improved the paper. We used \textsc{splash} for the figures and renderings \citep{price07}.

\bibliography{dan}

\begin{thebibliography}{}

\bibitem[\protect\citeauthoryear{{Ayliffe}, {Laibe}, {Price} \&
  {Bate}}{{Ayliffe} et~al.}{2012}]{ayliffeetal12}
{Ayliffe} B.~A.,  {Laibe} G.,  {Price} D.~J.,    {Bate} M.~R.,  2012, \mnras,
  423, 1450

\bibitem[\protect\citeauthoryear{{Barranco}}{{Barranco}}{2009}]{barranco09}
{Barranco} J.~A.,  2009, \apj, 691, 907

\bibitem[\protect\citeauthoryear{{Brookshaw}}{{Brookshaw}}{1985}]{brookshaw85}
{Brookshaw} L.,  1985, PASA, 6, 207

\bibitem[\protect\citeauthoryear{{Chiang}}{{Chiang}}{2008}]{chiang08}
{Chiang} E.,  2008, \apj, 675, 1549

\bibitem[\protect\citeauthoryear{{Cleary} \& {Monaghan}}{{Cleary} \&
  {Monaghan}}{1999}]{clearymonaghan99}
{Cleary} P.~W.,  {Monaghan} J.~J.,  1999, J. Comp. Phys., 148, 227

\bibitem[\protect\citeauthoryear{{Cullen} \& {Dehnen}}{{Cullen} \&
  {Dehnen}}{2010}]{cullendehnen10}
{Cullen} L.,  {Dehnen} W.,  2010, \mnras, 408, 669

\bibitem[\protect\citeauthoryear{{Espa{\~n}ol} \& {Revenga}}{{Espa{\~n}ol} \&
  {Revenga}}{2003}]{espanolrevenga03}
{Espa{\~n}ol} P.,  {Revenga} M.,  2003, \pre, 67, 026705

\bibitem[\protect\citeauthoryear{{Flebbe}, {Muenzel}, {Herold}, {Riffert} \&
  {Ruder}}{{Flebbe} et~al.}{1994}]{flebbeetal94}
{Flebbe} O.,  {Muenzel} S.,  {Herold} H.,  {Riffert} H.,    {Ruder} H.,  1994,
  \apj, 431, 754

\bibitem[\protect\citeauthoryear{{Jacquet}, {Balbus} \& {Latter}}{{Jacquet}
  et~al.}{2011}]{jbl11}
{Jacquet} E.,  {Balbus} S.,    {Latter} H.,  2011, \mnras, 415, 3591

\bibitem[\protect\citeauthoryear{{Laibe}, {Gonzalez} \& {Maddison}}{{Laibe}
  et~al.}{2012}]{LGM12}
{Laibe} G.,  {Gonzalez} J.-F.,    {Maddison} S.~T.,  2012, \aap, 537, A61

\bibitem[\protect\citeauthoryear{{Laibe} \& {Price}}{{Laibe} \&
  {Price}}{2011}]{laibeprice11}
{Laibe} G.,  {Price} D.~J.,  2011, \mnras, 418, 1491

\bibitem[\protect\citeauthoryear{{Laibe} \& {Price}}{{Laibe} \&
  {Price}}{2012a}]{laibeprice12}
{Laibe} G.,  {Price} D.~J.,  2012a, \mnras, 420, 2345

\bibitem[\protect\citeauthoryear{{Laibe} \& {Price}}{{Laibe} \&
  {Price}}{2012b}]{laibeprice12a}
{Laibe} G.,  {Price} D.~J.,  2012b, \mnras, 420, 2365

\bibitem[\protect\citeauthoryear{{Laibe} \& {Price}}{{Laibe} \&
  {Price}}{2014a}]{laibeprice14}
{Laibe} G.,  {Price} D.~J.,  2014a, \mnras, 440, 2136

\bibitem[\protect\citeauthoryear{{Laibe} \& {Price}}{{Laibe} \&
  {Price}}{2014b}]{laibeprice14a}
{Laibe} G.,  {Price} D.~J.,  2014b, \mnras, 440, 2147

\bibitem[\protect\citeauthoryear{{Laibe} \& {Price}}{{Laibe} \&
  {Price}}{2014c}]{laibeprice14b}
{Laibe} G.,  {Price} D.~J.,  2014c, \mnras, 444, 1940

\bibitem[\protect\citeauthoryear{{Lee}, {Chiang}, {Asay-Davis} \&
  {Barranco}}{{Lee} et~al.}{2010}]{leeetal10}
{Lee} A.~T.,  {Chiang} E.,  {Asay-Davis} X.,    {Barranco} J.,  2010, \apj,
  718, 1367

\bibitem[\protect\citeauthoryear{{Lodato} \& {Price}}{{Lodato} \&
  {Price}}{2010}]{lodatoprice10}
{Lodato} G.,  {Price} D.~J.,  2010, \mnras, 405, 1212

\bibitem[\protect\citeauthoryear{{Lor{\'e}n-Aguilar} \&
  {Bate}}{{Lor{\'e}n-Aguilar} \& {Bate}}{2014}]{lorenbate14}
{Lor{\'e}n-Aguilar} P.,  {Bate} M.~R.,  2014, \mnras, 443, 927

\bibitem[\protect\citeauthoryear{{Miura} \& {Glass}}{{Miura} \&
  {Glass}}{1982}]{miuraglass82}
{Miura} H.,  {Glass} I.~I.,  1982, Royal Society of London Proceedings Series
  A, 382, 373

\bibitem[\protect\citeauthoryear{{Monaghan}}{{Monaghan}}{2005}]{monaghan05}
{Monaghan} J.~J.,  2005, Reports on Progress in Physics, 68, 1703

\bibitem[\protect\citeauthoryear{{Morris} \& {Monaghan}}{{Morris} \&
  {Monaghan}}{1997}]{morrismonaghan97}
{Morris} J.~P.,  {Monaghan} J.~J.,  1997, J. Comp. Phys., 136, 41

\bibitem[\protect\citeauthoryear{{Pandey} \& {Wardle}}{{Pandey} \&
  {Wardle}}{2008}]{pandeywardle08}
{Pandey} B.~P.,  {Wardle} M.,  2008, \mnras, 385, 2269

\bibitem[\protect\citeauthoryear{{Price}}{{Price}}{2004}]{price04}
{Price} D.~J.,  2004, PhD thesis, University of Cambridge, Cambridge, UK.
  astro-ph/0507472

\bibitem[\protect\citeauthoryear{{Price}}{{Price}}{2007}]{price07}
{Price} D.~J.,  2007, \pasa, 24, 159

\bibitem[\protect\citeauthoryear{{Price}}{{Price}}{2008}]{price08}
{Price} D.~J.,  2008, J. Comp. Phys., 227, 10040

\bibitem[\protect\citeauthoryear{{Price}}{{Price}}{2012}]{price12}
{Price} D.~J.,  2012, J. Comp. Phys., 231, 759

\bibitem[\protect\citeauthoryear{{Price} \& {Federrath}}{{Price} \&
  {Federrath}}{2010}]{pricefederrath10}
{Price} D.~J.,  {Federrath} C.,  2010, \mnras, 406, 1659

\bibitem[\protect\citeauthoryear{{Price} \& {Monaghan}}{{Price} \&
  {Monaghan}}{2004}]{pricemonaghan04a}
{Price} D.~J.,  {Monaghan} J.~J.,  2004, \mnras, 348, 139

\bibitem[\protect\citeauthoryear{{Price} \& {Monaghan}}{{Price} \&
  {Monaghan}}{2007}]{pricemonaghan07}
{Price} D.~J.,  {Monaghan} J.~J.,  2007, \mnras, 374, 1347

\bibitem[\protect\citeauthoryear{{Sod}}{{Sod}}{1978}]{sod78}
{Sod} G.~A.,  1978, J. Comp. Phys., 27, 1

\bibitem[\protect\citeauthoryear{{Wadsley}, {Veeravalli} \&
  {Couchman}}{{Wadsley} et~al.}{2008}]{wvc08}
{Wadsley} J.~W.,  {Veeravalli} G.,    {Couchman} H.~M.~P.,  2008, \mnras, 387,
  427

\bibitem[\protect\citeauthoryear{{Watkins}, {Bhattal}, {Francis}, {Turner} \&
  {Whitworth}}{{Watkins} et~al.}{1996}]{watkinsetal96}
{Watkins} S.~J.,  {Bhattal} A.~S.,  {Francis} N.,  {Turner} J.~A.,
  {Whitworth} A.~P.,  1996, \aaps, 119, 177

\bibitem[\protect\citeauthoryear{{Youdin} \& {Goodman}}{{Youdin} \&
  {Goodman}}{2005}]{youdingoodman05}
{Youdin} A.~N.,  {Goodman} J.,  2005, \apj, 620, 459

\end{thebibliography}

\appendix
\begin{onecolumn}
\section{Proof that Equation~\ref{EQ:SPHDUDTALT} is a discrete form of Equation~\ref{eq:dudt}}
\label{sec:dudtalt}
 Here, we prove that the expression obtained for the second term in Eq.~\ref{EQ:SPHDUDTALT} by enforcing the conservation of energy, namely
\begin{equation}
-\frac{1}{2(1 - \epsilon_{a})\rho_{a}}\sum_{b} \frac{m_{b}}{\rho_{b}} (u_{a} - u_{b}) (D_{a} + D_{b}) (P_{a} - P_{b}) \frac{F_{ab}}{\vert r_{ab} \vert},
\label{eq:weird}
\end{equation}
is indeed a discrete form of the corresponding term in Eq.~\ref{eq:dudt}, i.e.
\begin{equation}
\frac{\epsilon\ts}{\rho_{\rm g}} \nabla P \cdot\nabla u.
\end{equation}
We proceed, following \citet{price12}, by identifying $-2 F_{ab}/\vert r_{ab}\vert$ as equivalent to the second derivative of a (new) kernel function, i.e. 
\begin{equation}
\nabla^{2} Y_{ab} \equiv \frac{-2 F_{ab}}{\vert r_{ab}\vert}.
\end{equation}
It may be shown straightforwardly that this new kernel $Y_{ab}$ indeed satisfies the normalisation conditions appropriate to the kernel second derivative (see \citealt{price12} for more details). We can then take the Laplacian of the standard SPH summation interpolant with this kernel, i.e.
\begin{equation}
A_{a} \simeq \sum_{b} m_{b} \frac{A_{b}}{\rho_{b}} Y_{ab},
\end{equation}
to give
\begin{equation}
\nabla^{2} A_{a} \simeq \sum_{b} m_{b} \frac{A_{b}}{\rho_{b}} \nabla^{2} Y_{ab}.
\label{eq:del2y}
\end{equation}
By writing (\ref{eq:weird}) in the form
\begin{equation}
\frac{1}{4\rho^{a}_{\rm g}} \sum_{b} \frac{m_{b}}{\rho_{b}} (u_{a} - u_{b}) (D_{a} + D_{b}) (P_{a} - P_{b}) \nabla^{2} Y_{ab},
\label{eq:weirdy}
\end{equation}
we can then use (\ref{eq:del2y}) to translate the various terms. Expanding (\ref{eq:weirdy}) we have
\begin{equation}
\frac{1}{4\rho^{a}_{\rm g}} \sum_{b} \frac{m_{b}}{\rho_{b}} (P_{a} u_{a} D_{a} - P_{a} u_{b} D_{a} +  P_{a} u_{a} D_{b} - P_{a} u_{b} D_{b} - P_{b} u_{a} D_{a} + P_{b} u_{b} D_{a} - P_{b}u_{a} D_{b} + P_{b} u_{b} D_{b} )\nabla^{2} Y_{ab}.
\end{equation}
Translating each of the terms in turn using (\ref{eq:del2y}) gives
\begin{equation}
\frac{1}{4\rho_{\rm g}}\left [P u D \nabla^{2} 1  -P D \nabla^{2} u + P u \nabla^{2} D - P \nabla^{2} (u D) - uD \nabla^{2} P + D \nabla^{2} (Pu)  - u \nabla^{2} (PD) + \nabla^{2} (PuD) \right].
\label{eq:terms}
\end{equation}
Expanding the $\nabla^{2} (ab)$ terms using the vector identity
\begin{equation}
\nabla^{2} (ab) = a \nabla^{2} b + 2( \nabla a \cdot \nabla b) + b \nabla^{2} a,
\end{equation}
and expanding the last term using
\begin{equation}
\nabla^{2} (PuD) = uD \nabla^{2} P + PD\nabla^{2} u + Pu \nabla^{2} D + 2u (\nabla P\cdot\nabla D) + 2D(\nabla P\cdot\nabla u) + 2P(\nabla D \cdot \nabla u),
\end{equation}
we find, upon simplification that (\ref{eq:terms}) reduces to simply
\begin{equation}
\frac{1}{4\rho_{\rm g}}\left [ 4D (\nabla P \cdot \nabla u)\right].
\end{equation}
Hence, (\ref{eq:weirdy}) and so (\ref{eq:weird}) is a discrete form of
\begin{equation}
 \frac{D}{\rho_{\rm g}} (\nabla P \cdot \nabla u) = \frac{\epsilon \ts}{\rho_{\rm g}} \nabla P \cdot \nabla u.
\end{equation}
QED.

\section{Enforcing positivity of the dust fraction}
\label{sec:s}
 While the usual SPH density summation enforces positivity of the total density, in the one fluid approach there is no constraint on the positivity of the dust fraction, being simply evolved via a differential equation. We found during our testing of the algorithm (Section~\ref{sec:settle}) that this can occur in practice, even though we conserve the total dust mass. An simple example is where $\epsilon$ is non-zero on only a fraction of the particles and zero on others, implying an infinite gradient in $\epsilon$ at the discontinuity surface, which as the dust front evolves can lead to negative $\epsilon$ on the particles that initially had zero. It should be noted that however that those errors are small and kernel dependant (i.e. of order $10^{-5}$ with a quintic kernel).  We discuss other possible solutions in Section~\ref{sec:settle}, but here present one such solution, which is to evolve the quantity 
\begin{equation}
s = \sqrt{\rho \epsilon},
\end{equation}
instead of $\epsilon$. We can enforce the same conservation of dust mass but with a guaranteed positivity of the dust fraction since
\begin{equation}
\epsilon_{a} = s_{a}^{2}/\rho_{a}.
\end{equation}
 
\subsection{Continuum equation}
  In terms of $s$, the local equation for dust mass conservation is
\begin{align}
\frac{\mathrm{d} s}{\mathrm{d}t} & = -\frac{\rho}{2s} \nabla \cdot \left[s^{2} \left(1 - \frac{s^{2}}{\rho} \right) \Delta\bm{v} \right] - \frac{s}{2} \nabla \cdot \bm{v}, \nonumber \\
& = -\frac{\rho}{2} \left\{ \nabla \cdot \left[ s\left(1 - \frac{s^{2}}{\rho} \right) \Delta\bm{v} \right]  + \left(1 - \frac{s^{2}}{\rho} \right) \Delta\bm{v} \cdot \nabla s \right\} - \frac{s}{2} \nabla \cdot \bm{v}.
\label{eq:dsdt}
\end{align}
where as previously ${\mathrm{d} }/{\mathrm{d}t}$ is the convective derivative using the barycentric velocity. For the case of hydrodynamics (Section~\ref{sec:hydro}) where $\Delta \bm{v} = t_{\rm s} \nabla P / [(1 - s^{2}/\rho) \rho]$, we can simplify this to\begin{equation}
\frac{\mathrm{d} s}{\mathrm{d}t} =  -\frac{\rho}{2} \left[\nabla\cdot \left( \frac{s t_{\rm s} \nabla P}{\rho}\right) + \frac{t_{\rm s}}{\rho} \nabla P \cdot\nabla s \right] - \frac{s}{2} \nabla \cdot \bm{v}.
\label{eq:dsdthydro}
\end{equation}

\subsection{SPH implementation with two first derivatives}
 Conservation of the total dust mass implies
\begin{equation}
\frac{\mathrm d}{\mathrm{d} t} \left(\sum_{a} \frac{m_{a}}{\rho_{a}} s_{a}^{2} \right) =0 ,
\end{equation}
giving
\begin{equation}
2 \sum_{a} \frac{m_{a}}{\rho_{a}} s_{a} \frac{\mathrm d s_{a}}{\mathrm{d} t} = \sum_{a} \frac{m_{a}}{\rho_{a}^{2}} \frac{\mathrm d \rho_{a}}{\mathrm{d} t} s_{a}^{2} .
\label{eq:cons_sa_SPH}
\end{equation} 
To enforce Eq.~\ref{eq:cons_sa_SPH}, we compute the evolution of $s_{a}$ according to
\begin{equation}
\frac{\mathrm{d} s_{a}}{\mathrm{d}t} = -\frac{\rho_{a}}{2} \sum_{b}m_{b}s_{b} \left( \frac{1 -s_{a}^{2}/\rho_{a}}{\Omega_{a}\rho_{a}^{2}} \Delta \bm{v}_{a} \cdot \nabla_{a} W_{ab} \left(h_{a} \right) + \frac{1 -s_{b}^{2}/\rho_{b}}{\Omega_{b}\rho_{b}^{2}} \Delta \bm{v}_{b} \cdot \nabla_{b} W_{ab} \left(h_{b} \right) \right) + \frac{s_{a}}{2\rho_{a}\Omega_{a}} \sum_{b}m_{b} \left( \bm{v}_{a} - \bm{v}_{b}\right) \cdot  \nabla_{a} W_{ab}\left(h_{a} \right) .
\label{eq:dsadt}
\end{equation}
The first term of the right-hand side of Eq.~\ref{eq:dsadt} corresponds to the first two terms (i.e. inside the brackets) of the right-hand side of Eq.~\ref{eq:dsdt} in the continuous limit (note the factor $s_{b}$ \textit{inside} the SPH summation). The contribution of this term to the left hand-side of Eq.~\ref{eq:cons_sa_SPH} is zero as it leads to a double summation of an antisymmetric term with respect to the indices $a$ and $b$. The second term of the right-hand side of Eq.~\ref{eq:dsadt} corresponds to the $\nabla \cdot \bm{v}$ term of Eq.~\ref{eq:dsdt} and provides the right hand-side of Eq.~\ref{eq:cons_sa_SPH}, which can be seen by differentiating the SPH density summation (\ref{eq:sphcty}) with respect to time.

\subsection{SPH implementation with direct second derivatives}
 We can also construct a method evolving $s$ but with direct second derivatives. Here we discretise  Eq.~\ref{eq:dsdthydro} using
\begin{equation}
\frac{\mathrm{d} s_{a}}{\mathrm{d}t} = -\frac{\rho_{a}}{2} \sum_{b} \frac{m_{b} s_{b}}{\overline{\rho}_{ab}} (D_{a} + D_{b}) (P_{a} - P_{b}) \frac{\overline{F}_{ab}}{\vert r_{ab}\vert} + \frac{s_{a}}{2\rho_{a}\Omega_{a}} \sum_{b} m_{b} \bm{v}_{ab}\cdot \nabla W_{ab} (h_{a}),
\label{eq:dsdtsph2}
\end{equation}
where $D \equiv s^{2} t_{\rm s}/\rho$ as previously. It is straightforward to show that this is indeed a discretisation of (\ref{eq:dsdthydro}) using the method described in Appendix~\ref{sec:dudtalt}. The average density is required in the denominator in order to conserve the total dust mass, which can be verified by substituting (\ref{eq:dsdtsph2}) into (\ref{eq:cons_sa_SPH}).

\end{onecolumn}

\label{lastpage}
\end{document}